\documentclass[aps,prd,nofootinbib,preprintnumbers,eqsecnum]{revtex4}
\usepackage{amsmath}
\usepackage{amssymb, amsmath}
\usepackage{amsfonts}
\usepackage{graphicx}
\usepackage{bm}

\usepackage{color}

\def\mc{\mathcal}
\def\mr{\mathrm}
\def\pd{\partial}

\newcommand{\diff}[3]{\frac{\mr{d}^{#1} #2}{\mr{d}#3^{#1}}}

\begin{document}

\preprint{YITP-13-120}

\title{Coleman-de Luccia instanton in dRGT massive gravity}

\date{\today}
\author{Ying-li Zhang\footnote{E-mail
 address: yingli{}@{}yukawa.kyoto-u.ac.jp}, Ryo Saito\footnote{E-mail
 address: rsaito{}@{}yukawa.kyoto-u.ac.jp}, Dong-han Yeom\footnote{E-mail
 address: innocent.yeom{}@{}gmail.com}
  and Misao Sasaki\footnote{E-mail
 address: misao{}@{}yukawa.kyoto-u.ac.jp}
 }
\affiliation{Yukawa Institute for Theoretical Physics, Kyoto
University, Kyoto 606-8502, Japan }

\begin{abstract}
We study the Coleman-de Luccia (CDL) instanton characterizing the
tunneling from a false vacuum to the true vacuum in a semi-classical
way in dRGT (deRham-Gabadadze-Tolley) massive gravity theory, and
evaluate the dependence of the tunneling rate on the model
parameters. It is found that provided with the same physical Hubble
parameters for the true vacuum $H_{\rm T}$ and the false vacuum
$H_{\rm F}$ as in General Relativity (GR), the thin-wall
approximation method implies the same tunneling rate as GR. However,
deviations of tunneling rate from GR arise when one goes beyond the
thin-wall approximation and they change monotonically until the
Hawking-Moss (HM) case. Moreover, under the thin-wall approximation,
the HM process may dominate over the CDL one if the value for the
graviton mass is larger than the inverse of the radius of the
bubble.
%Moreover, under the thin-wall approximation, it is found that for
%$Y_\pm>0$, HM process may dominate over CDL one in varying-mass
%scenario or with Minkowskian fiducial metric.
\end{abstract}

\maketitle

%%%%%%%%%%%%%%%%%%%%%%%%%
\section{Introduction}
%%%%%%%%%%%%%%%%%%%%%%%%%

The notion of gravitational theory with a massive graviton is
nothing new. A massive gravity theory was firstly proposed by Fierz
and Pauli (FP) in 1939 where the theory of General Relativity (GR)
was extended by a linear mass term~\cite{Fierz:1939}. However, %since
%relativistic and non-relativistic matter couple to the gravity
%sector with different coupling strength regardless of the value of
%mass of graviton, the FP theory suffers from the van
%Dam-Veltman-Zakharov (vDVZ) discontinuity
%problem~\cite{vDV:1970,Zakharov:1970}. Although Vainshtein mechanism
%helps to relieve this problem by introducing nonlinear
%terms~\cite{Vainshtein:1972},
lack of Hamiltonian and momentum constraints leads to 6 degrees of
freedom in this theory, with 5 of which corresponding to those of a
massive spin-2 graviton and the rest one a Boulware-Deser (BD) ghost
mode~\cite{Boulware:1972,Creminelli:2005qk,Rubakov:2008,Hinterbichler:2012}.
A breakthrough was achieved by recent development of de
Rham-Gabadadze-Tolley (dRGT) nonlinear massive gravity
theory~\cite{Rham:2010,Rham:2011PRL,Hassan:2011vm}, where a special
form of potential was introduced to recover the Hamiltonian
constraint so that the sixth BD ghost mode is
eliminated~\cite{Hassan:2012,Hassan:2011tf}. One of the most
remarkable consequences in this theory is that it allows
self-accelerating solutions~\cite{D'Amico:2011jj,
Gumrukcuoglu:2011open, Kobayashi:2012, Gratia:2012}, where the
universe takes the de Sitter form even without a bare cosmological
constant, and its Hubble scale is of the order of the graviton mass.

However, application of the self-accelerating solution to explain
the current accelerated expansion of universe does not solve the
``Cosmological Constant Problem''
(CCP)~\cite{Weinberg:1988cp,Nobbenhuis:2004wn}, which implies a
serious contradiction between smallness of the cosmological constant
and expected large quantum corrections. Motivated by the proposal
that CCP may hopefully be solved by the anthropic selection of the
cosmological constant in the landscape of
vacua~\cite{Weinberg:1988cp, Susskind:2003kw, Nobbenhuis:2004wn},
the Hawking-Moss (HM) solution~\cite{Hawking:1981fz} in dRGT massive
gravity was studied in~\cite{ZSS:2013}. It was found that depending
on the choice of parameters in dRGT massive gravity theory, the
non-vanishing mass of a graviton will influence the tunneling rate
of HM instanton, hence affects the stability of a vacuum in this
theory.

On the other hand, it was known that in GR, traditionally, the
Hartle-Hawking (HH) no-boundary wavefunction~\cite{Hartle:1983}
exponentially prefers small number of $e$-foldings near the minimum
of the inflaton potential, hence it does not seem to predict the
universe we observe today. Such situation may be drastically changed
by application of the correction term in the Euclidean HM action to
HH no-boundary proposal. It was found that for a wide range choice
of the parameters in dRGT massive gravity theory, the no-boundary
wavefunction can peak at a sufficiently large value of the Hubble
parameter, hence one may obtain a sufficient number of $e$-folds of
inflation~\cite{SYZ:2013}.

Inspired by these interesting achievements, as a necessary step
towards understanding of tunneling process in dRGT massive gravity
theory, it is necessary to explore another kind of phase transition
in theories of scalar fields coupled to gravity: Coleman-de Luccia
(CDL) instanton that exists for a special form of potentials such
that the curvature scale of the barrier is large compared to the
potential energy (in Planck units)~\cite{Coleman:1980,BK:2006}, and
gradually approaches to the HM instanton as the curvature scale
shrinks. In this paper, we consider the CDL solution for a scalar
field with minimal coupling to gravity in dRGT massive gravity
theory. We set up the model and found the bounce solutions
corresponding to the CDL instantons. Based on these solutions, we
evaluate the CDL action in two approaches: ``thin-wall''
approximation and perturbations around HM solution, and find
monotonic contributions from the graviton mass terms until the
``thick-wall'' limit: the HM case. Moreover, in the ``thin-wall''
limit, comparison of the tunneling rates for the HM and CDL
solutions with the same potential shows that, even when the CDL
process dominates over the HM one in the case of GR, it may behave
inversely in the context of dRGT massive gravity, i.e. depending on
the values of parameters in this model, the HM process may dominate
over the CDL one.

This paper is organized as follows. In Sec.~\ref{setup}, we setup
the Lagrangian for our model. In Sec.~\ref{s:basic}, we formulate
the equations of motion (EOM) and solve the constraint equation. In
Sec. \ref{s:CDLthin}, the CDL solution is studied by using
``thin-wall'' approximation. In Sec.~\ref{s:CDLperturb}, we study
the CDL solution as perturbations around the Hawking-Moss (HM)
solution obtained in Ref.~\cite{ZSS:2013} and clarify its
relationship with the one obtained by ``thin-wall'' approximation.
In Sec.~\ref{s:HMvsCDL}, we compare the tunneling rates for the CDL
and HM instantons for the same potential which satisfies the
condition for ``thin-wall'' approximation. In Sec.~\ref{conclusion},
we draw the conclusions. In Appendix.~\ref{perturbHM}, we present
the detailed calculations of the CDL process as perturbations around
HM one. Appendix.~\ref{cdltoHM} is devoted to the deduction of
comparison of the CDL to HM probabilities.

Throughout the paper, the Lorentzian metric signature is set to be
$(-,+,+,+)$, while the Euclidean metric signature $(+,+,+,+)$.
Meanwhile, we use the conventional notations where indices with
Greek letters $\mu, \nu,...$ for the spacetime indices, the Latin
letters $i,j,...$ for the space indices, while the Latin indices
$a,b,...$ for the internal space (Lorentz frame) indices. Also
repeated indices imply the summation unless otherwise stated.

%%%%%%%%%%%%%%%%%%%%%%%%%
\section{Setup of model}\label{setup}
%%%%%%%%%%%%%%%%%%%%%%%%%

\begin{figure}
\includegraphics[height=6.5cm,keepaspectratio=true,angle=0]{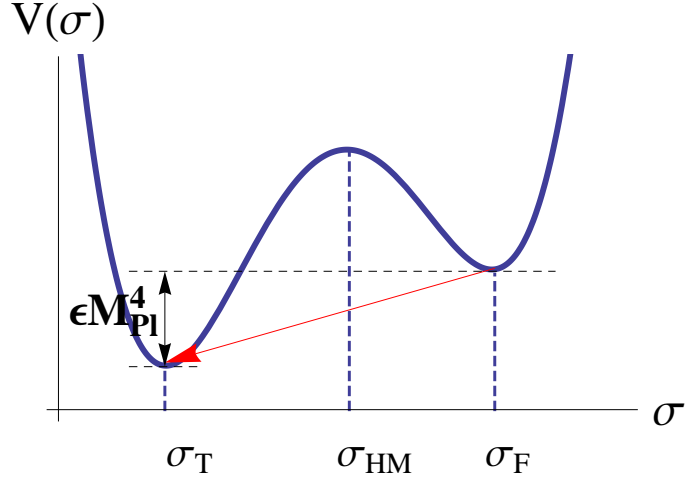} \caption{Illustration of the potential $V(\sigma)$ with two
local minima $\sigma_{\rm F}$ and $\sigma_{\rm T}$, which correspond
to the false and true vacuum respectively, while $\sigma_{\rm HM}$
labels its local maximum value. In the thin-wall approximation, the
difference of the potential values at these two local minima is very
small, i.e. $V(\sigma_{\rm F})-V(\sigma_{\rm T})=\epsilon\ll1$.}
 \label{fig:potential}
\end{figure}

We study the tunneling process of a minimally coupled scalar field
$\sigma$ which tunnels from one vacua $\sigma_{\rm F}$ to another
one $\sigma_{\rm T}$ in the context of dRGT massive gravity, as
illustrated in Fig.~\ref{fig:potential}. The dRGT massive gravity is
composed of two metrics, namely a physical metric $g_{\mu\nu}$ and a
fiducial
 metric $G_{ab}$, with the St\"{u}ckelberg fields $\phi^a$~\cite{Rham:2010,Rham:2011PRL}. As usual, the whole action can be
divided into two parts: the nonlinear massive gravity part $I_g$ and
the minimally coupled scalar field part $I_m$ as follows:
\footnote{It should be noted that we use the natural units where
$M_{\rm Pl}^{-2}\equiv8\pi G=1$ throughout this paper.}
\begin{align}\label{eq:actionorig}
  S &= I_g+I_m, \\
  \label{eq:actiong}I_g &\equiv \int\mr{d}^4x~\sqrt{-g}\left[ \frac{R}{2}
 + m_g^2(\mc{L}_2 + \alpha_3\mc{L}_3 + \alpha_4\mc{L}_4)\right],
\\
  I_m &\equiv -\int\mr{d}^4x~\sqrt{-g}\left[ \frac{1}{2}(\pd \sigma)^2
 + V(\sigma) \right],
\end{align}
where $m_g$, $\alpha_3$ and $\alpha_4$ are three free parameters in
this model, and
\begin{align}\label{eq:mass}
   {\cal L}_2 &= \frac{1}{2}\left(\left[{\cal K}\right]^2
-\left[{\cal K}^2\right]\right)\,,
\nonumber\\
   {\cal L}_3 &= \frac{1}{6}\left(\left[{\cal K}\right]^3
-3\left[{\cal K}\right]\left[{\cal K}^2\right]+2\left[{\cal K}^3\right]\right),
\nonumber\\
  {\cal L}_4 & = \frac{1}{24} \left(\left[{\cal K}\right]^4
-6\left[{\cal K}\right]^2\left[{\cal K}^2\right]
+3\left[{\cal K}^2\right]^2+8\left[{\cal K}\right]\left[{\cal K}^3\right]
-6\left[{\cal K}^4\right]\right),
\end{align}
with
\begin{align}\label{eq:k}
   {\cal K}^{\mu}_{\nu} \equiv \delta^{\mu}_{\nu}
 - \sqrt{g^{\mu\sigma}G_{ab}(\phi)\pd_{\nu}\phi^a\pd_{\sigma}\phi^b}.
    \end{align}

In order to investigate tunneling process, the potential $V(\sigma)$
is assumed to have two local minima $\sigma_{\rm F}$ and
$\sigma_{\rm T}$ which correspond to the false and true vacuum,
respectively, with a local maximum between them, $\sigma=\sigma_{\rm
HM}$, as illustrated in Fig. \ref{fig:potential}.

The Euclidean action of (\ref{eq:actionorig}) is obtained by Wick
rotation $x^0 \to ix_{\rm E}^0$ and correspondingly $S_{\rm E}=iS$.
In the semiclassical limit, the tunneling rate per unit time per
unit volume can be expressed in terms of the Euclidean action as
follows:
\begin{align}
    P\equiv\Gamma/V= Ae^{-B}\,;
\quad
    B = S_{\rm E}[{g}_{\mu\nu,{\rm B}},{\phi}_{\rm B}]
-S_{\rm E}[{g}_{\mu\nu,{\rm F}},{\phi}_{\rm F}]\,, \label{eq:rate}
\end{align}
where $\{ {g}_{\mu\nu,{\rm B}}, {\phi}_{\rm B} \}$ is the bounce
solution, which is a solution of the Euclidean equations of motion
with appropriate boundary conditions, and $\{ {g}_{\mu\nu,{\rm F}},
{\phi}_{\rm F} \}$ is the solution of false
vacuum~\cite{Coleman:1980}. Conventionally, the bounce solution $\{
{g}_{\mu\nu,{\rm B}}, {\phi}_{\rm B} \}$ is explored under the
assumption of $O(4)$-symmetry, since an $O(4)$-symmetric solution
gives the lowest action for a wide class of scalar-field theories,
hence gives the least value of action which dominates the tunneling
process~\cite{Coleman:1977th,Lee:2008,Lee:2009,Lee:2011,Lee:2013}.
The same assumption is also reasonable in the presence of gravity
\cite{Coleman:1980}, therefore, the physical metric can be assumed
to take the following form,
 \begin{align}\label{eq:o4metric}
        g_{\mu\nu}\mr{d}x^{\mu}\mr{d}x^{\nu} = N(\xi)^2\mr{d}\xi^2
 + a(\xi)^2\Omega_{ij}\mr{d}x^i\mr{d}x^j,
    \end{align}
where $\Omega_{ij}\mr{d}x^i\mr{d}x^j$ is
 the metric on a three-sphere with $K>0$,
   \begin{align}\label{eq:3sphere}
           \Omega_{ij} &\equiv \delta_{ij}
 + \frac{K\delta_{il}\delta_{jm}x^l x^m}{1-K\delta_{lm}x^l x^m}\,.
    \end{align}

In the context of dRGT massive gravity, the fiducial metric $G_{ab}$
is assumed to be non-dynamical~\cite{Hassan:2011zd, Volkov:2011an,
Comelli:2011zm}. In order to guarantee the $O(4)$-symmetry, we
assume that it is given by the de Sitter
metric~\cite{Langlois:2012hk} with a constant Hubble parameter
$F$~\cite{ZSS:2013}:
\begin{align}\label{eq:reference}
 G_{ab}(\phi)\mr{d}\phi^a\mr{d}\phi^b
&\equiv -(\mr{d}\phi^0)^2 +b(\phi^0)^2\Omega_{ij} \mr{d}\phi^i\mr{d}\phi^j,
\end{align}
where
\begin{equation} \label{eq:bdefinition}
\left\{ \begin{aligned}
         b(\phi^0) &\equiv F^{-1}\sqrt{K}\cosh(F\phi^0)\,, \\
         \phi^0 &= f(\xi)\,, \quad \phi^i = x^i\,.
                          \end{aligned} \right.
                          \end{equation}

It should be noted that here we stick to the Lorentzian signature
for the fiducial metric since it is non-dynamical in dRGT massive
gravity theory. Nevertheless, thanks to the assumption of the de
Sitter fiducial metric, we may adopt the $O(4)$-ansatz.

%%%%%%%%%%%%%%%%%%%%%%%%%
\section{Euclidean equations of motion} \label{s:basic}
%%%%%%%%%%%%%%%%%%%%%%%%%
In this section, we present the Euclidean equations of motion. More
detailed deductions can be found in~\cite{ZSS:2013}.

Inserting the $O(4)$-ansatz (\ref{eq:o4metric}) and
(\ref{eq:bdefinition}) into (\ref{eq:k}), one obtains the Euclidean
version of the action (\ref{eq:action})

\begin{align}\label{eq:action}
  S_{\rm E} &= I_{g {\rm E}}+I_{m {\rm E}},
\end{align}
where the gravity action is reduced to
\begin{align}\label{eq:euclidean}
        I_{g{\rm E}} = \int\mr{d}^4x_E~\sqrt{\Omega}
\left[-3KNa - \frac{3\dot{a}^2 a}{N}
 - m_g^2\left(\mathcal{L}_{2E}+\alpha_3\mathcal{L}_{3E}+\alpha_4\mathcal{L}_{4E}\right)\right],
\end{align}
with
    \begin{subequations}\label{eq:mass}
     \begin{align}
       \mathcal{L}_{2E} &= 3a(a-b)\left(2Na-\sqrt{-\dot{f}^2}a-Nb\right),\\
        \mathcal{L}_{3E} &= (a-b)^2\left(4Na-3\sqrt{-\dot{f}^2}a-Nb\right), \\
        \mathcal{L}_{4E} &= (a-b)^3\left(N-\sqrt{-\dot{f}^2}\right),
         \end{align}
     \end{subequations}
and a dot means a derivative with respect to the radial coordinate,
$\dot{~} \equiv \mr{d}/\mr{d}\xi$. Meanwhile, the action for the
tunneling field is reduced to
    \begin{align}\label{eq:ematter}
        I_{mE} = \int\mr{d}^4x~a^3
\sqrt{\Omega}\left[\frac{1}{2N}\dot{\sigma}^2 + NV(\sigma)\right].
    \end{align}

Variation of the action (\ref{eq:euclidean}) with respect to the
St\"{u}ckelberg field $\phi^0=f$ gives the following constraint
equation:
\begin{align}\label{eq:EOM1}
        (i\dot{a}+Nb_{,f})\left[\left(3-\frac{2b}{a}\right)
 + \alpha_3\left(1-\frac{b}{a}\right)\left(3-\frac{b}{a}\right)
 + \alpha_4\left(1-\frac{b}{a}\right)^2\right] = 0\,,
\end{align}
where $b_{,f} \equiv {\rm d}b/{\rm d}f= \sqrt{K}\sinh(Ff)$.
Correspondingly, we obtain two branches:

\begin{align}
&{\rm Branch~I}~\qquad\qquad Nb_{,f} = -i\dot{a}\,,\label{eq:b1}\\
&{\rm Branch~II}~\qquad  \left(3-\frac{2b}{a}\right)
 + \alpha_3\left(1-\frac{b}{a}\right)\left(3-\frac{b}{a}\right)
 + \alpha_4\left(1-\frac{b}{a}\right)^2 = 0\,.\label{eq:b2}
\end{align}

In Branch I, it is known that there exists a tension between the
Vainstein mechanism and the Higuchi bound~\cite{Rham:2012p}. This
situation is not improved even in the extended massive gravity
theories or bigravity theory. Hence, in the following, we mainly
concentrate on analysis of Branch II. The solution to
Eq.~(\ref{eq:b2}) is given by
    \begin{align}\label{eq:II}
 b = X_{\pm}a\,,
\qquad X_{\pm} \equiv
\frac{1+2\,\alpha_3+\alpha_4\pm\sqrt{1+\alpha_3+\alpha_3^2-\alpha_4}}
{\alpha_3+\alpha_4}\,,
    \end{align}
where it should be noted that we require $X_\pm>0$ for our interest.
Hereafter, for definiteness, the choice of $X_+$ is called Branch
II$_+$ while $X_-$ is called Branch II$_-$. On the other hand,
variations of the action (\ref{eq:action}) with respect to lapse
function $N$ and $\sigma$ give the ``Friedman equation'' and field
equation respectively:
\begin{align}
 &\frac{3}{a^2}\left(a^{\prime 2} - K\right)
=\frac{\sigma^{\prime 2}}{2} - V(\sigma) - \Lambda_{\pm}\,,
 \label{eq:eom}
 \\
 &\sigma^{\prime\prime} + 3\mathcal{H}\sigma^{\prime}
 -V_{,\sigma}(\sigma) = 0\,,
 \label{eq:eom1}
\end{align}
where we have introduced the proper radial coordinate $\tau \equiv
\int N{\rm d}\xi$ and a prime means derivative with respect to the
proper time: $~^{\prime}\equiv{\rm d}/{\rm d}\tau$,
$\mathcal{H}\equiv a^\prime/a$, while
\begin{align}\label{eq:lambda}
 \Lambda_\pm &\equiv-m_g^2\left(1-X_\pm\right)
\left[3\left(2-X_\pm\right) +
\alpha_3\left(1-X_\pm\right)\left(4-X_\pm\right)
 + \alpha_4\left(1-X_\pm\right)^2\right]\nonumber\\
 &=-\frac{m_g^2}{\left(\alpha_3+\alpha_4\right)^2}
\left[\left(1+\alpha_3\right)\left(2+\alpha_3
+2\,\alpha_3^2-3\,\alpha_4\right)
 \pm 2\,\left(1+\alpha_3 +\alpha_3^2-\alpha_4\right)^{3/2}\right]\,.
\end{align}

As explained in the previous section, the tunneling rate
(\ref{eq:rate}) is given by the Euclidean action evaluated for a
solution of Eqs.~(\ref{eq:eom}) and (\ref{eq:eom1}). In the
following, we construct the CDL solution and then evaluate the
action for it.

%%%%%%%%%%%%%%%%%%%%%%
\section{Coleman-DeLuccia solution from thin-wall approximation}\label{s:CDLthin}
%%%%%%%%%%%%%%%%%%%%%%

\subsection{Expression for Euclidean action}\label{s:CDL solution}
In order to evaluate the CDL solution by using thin-wall
approximation, one should firstly solve the Euclidean Friedmann
equation (\ref{eq:eom}) at the locally minimal point
$\sigma_{\rm{T}}$ and the globally minimal point $\sigma_{\rm{F}}$
separately:

\begin{equation} \label{eq:friedCDL}
\frac{3}{a^2}\left(a^{\prime 2} - K\right)=\left\{
\begin{aligned}
          & - V(\sigma_{\rm T}) - \Lambda_{\pm} \equiv -\Lambda_{\pm, \rm{T}}\,, \qquad \tau<\tau_0\\
          \\
          & - V(\sigma_{\rm F}) - \Lambda_{\pm} \equiv -\Lambda_{\pm,
          \rm{F}}\,, \qquad \tau>\tau_0
                          \end{aligned} \right.
                          \end{equation}
where $\tau_0$ is the point at which the tunneling process occurs.
Hence, from Eq.~(\ref{eq:friedCDL}), inside and outside solutions
can be obtained as

\begin{equation} \label{eq:aCDLsol}
a(\tau)\left\{ \begin{aligned}
         &=a_{\rm{T}}(\tau)\equiv H_{\rm{T}}^{-1}\sqrt{K}\cos\left(H_{\rm{T}}\tau+\theta_{\rm{T}}\right)\,,\qquad \tau<\tau_0 \\
         \\
         &=a_{\rm{F}}(\tau)\equiv H_{\rm{F}}^{-1}\sqrt{K}\cos\left(H_{\rm{F}}\tau+\theta_{\rm{F}}\right)\,, \qquad \tau>\tau_0
                          \end{aligned} \right.
                          \end{equation}
where for convenience, we set $\theta_{\rm{T}}=0$ in the following.
Moreover, we have introduced the inside/outside Hubble parameter of
the physical metric by
\begin{equation} \label{eq:hubblecdl}
\left\{ \begin{aligned}
         H_{\rm{T}} & \equiv  \sqrt{\frac{\Lambda_{\pm,{\rm T}}}{3}}
=  \sqrt{\frac{V(\sigma_{\rm T}) + \Lambda_{\pm}}{3}}\,,\\
         H_{\rm{F}} & \equiv  \sqrt{\frac{\Lambda_{\pm,{\rm F}}}{3}}
=  \sqrt{\frac{V(\sigma_{\rm F}) + \Lambda_{\pm}}{3}}\,,
                          \end{aligned} \right.
                          \end{equation}
whereas it should be noted that the phase angle $\theta_{\rm{F}}$ is
determined by the continuity condition on the shell $\tau=\tau_0$:
\begin{eqnarray}\label{eq:conti}
    H_{\rm{T}}^{-1}\cos\left(H_{\rm{T}}\tau_0\right)=H_{\rm{F}}^{-1}\cos\left(H_{\rm{F}}\tau_0+\theta_{\rm{F}}\right)\,.
\end{eqnarray}

Using the constraint equation (\ref{eq:II}) and the definition for
$b(\phi^0)$ Eq.~(\ref{eq:bdefinition}), the following relationship
holds:
\begin{eqnarray}\label{eq:bfa}
    b(\tau)\equiv F^{-1}\sqrt{K}\cosh\left(Ff(\tau)\right)=X_\pm
    a(\tau)\,,
\end{eqnarray}
from which the expression for $f(\tau)$ can be obtained as follows:

\begin{equation} \label{eq:fsol}
f(\tau)=\left\{ \begin{aligned}
          & F^{-1}\cosh^{-1}\left[\frac{FX_\pm}{H_{\rm{T}}}\cos\left(H_{\rm{T}}\tau\right)\right]\,, \qquad \tau<\tau_0 \\
          & F^{-1}\cosh^{-1}\left[\frac{FX_\pm}{H_{\rm{F}}}\cos\left(H_{\rm{F}}\tau+\theta_{\rm{F}}\right)\right]\,, \qquad \tau>\tau_0
                          \end{aligned} \right.
                          \end{equation}
hence, its derivative with respect to the proper radial coordinate
$\tau$ can be obtained as:
\begin{equation} \label{eq:fderi}
-\bigg(f^\prime(\tau)\bigg)^2=\left\{
\begin{aligned}
          & \frac{X_\pm^2\sin^2\left(H_{\rm{T}}\tau\right)}{\left(\frac{FX_\pm}{H_{\rm{T}}}\cos\left(H_{\rm{T}}\tau\right)\right)^2-1}=X_\pm^2\frac{K-\left(a_{\rm{T}}H_{\rm{T}}\right)^2}{K-\left(a_{\rm{T}}FX_\pm\right)^2}\,, \qquad \tau<\tau_0 \\
          & \frac{X_\pm^2\sin^2\left(H_{\rm{F}}\tau+\theta_{\rm F}\right)}{\left(\frac{FX_\pm}{H_{\rm{F}}}\cos\left(H_{\rm{F}}\tau+\theta_{\rm{F}}\right)\right)^2-1}=X_\pm^2\frac{K-\left(a_{\rm{F}}H_{\rm{F}}\right)^2}{K-\left(a_{\rm{F}}FX_\pm\right)^2}\,, \qquad \tau>\tau_0
                          \end{aligned} \right.
                          \end{equation}
where we note that provided with the continuity equation
(\ref{eq:conti}), $f(\tau)$ is continuous at $\tau=\tau_0$, but its
derivative ${\rm{d}}f(\tau)/{\rm{d}}\tau$ is discontinuous on the
shell.

Inserting Eqs.~(\ref{eq:aCDLsol}) and (\ref{eq:fderi}) into the
Euclidian action given by Eqs.~(\ref{eq:euclidean}) and
(\ref{eq:ematter}), and using $N^{-1}\dot f={\rm d}f/{\rm d}\tau$,
the total action can be expressed by division into three parts:
\begin{align}\label{eq:actionCDL}
S_{\rm E}[a(\tau), \sigma] &= \int {\rm d}^3x\sqrt{\Omega}
\int^{\pi/(2H_{\rm F})}_{-\pi/(2H_{\rm T})}{\rm d}\tau~ a^3(\tau)
\left[2\big(V(\sigma)+\Lambda_{\pm}\big)-\frac{6K}{a^2(\tau)}
 + m_g^2Y_\pm \sqrt{-(f^{\prime})^2}\right]\nonumber\\
 &=S_{\rm{inside}}+S_{\rm{outside}}+S_{\rm{wall}}\,,
\end{align}
where for brevity, we have introduced the parameter $Y_\pm$ in terms
of $X_{\pm}$ as follows:
\begin{eqnarray}\label{eq:Y}
    Y_\pm\equiv3(1-X_{\pm})+3\alpha_3(1-X_{\pm})^2+\alpha_4(1-X_{\pm})^3\,,
\end{eqnarray}
while $S_{\rm{inside}}$, $S_{\rm{outside}}$ and $S_{\rm{wall}}$ are
defined as:
\begin{align}
S_{\rm{inside}} &\equiv \int {\rm d}^3x\sqrt{\Omega}
\int^{\tau_0(1-\delta)}_{-\pi/(2H_{\rm T})}{\rm d}\tau~ a^3_{\rm T}
\left[2\Lambda_{\pm, {\rm T}}-\frac{6K}{a^2_{\rm T}}
 + m_g^2Y_\pm |X_\pm|\sqrt{\frac{K-\left(a_{\rm{T}}H_{\rm{T}}\right)^2}{K-\left(a_{\rm{T}}FX_\pm\right)^2}}\right]\,,\label{Sin}\\
S_{\rm{outside}} &\equiv \int {\rm d}^3x\sqrt{\Omega}
\int^{\pi/(2H_{\rm F})}_{\tau_0(1+\delta)}{\rm d}\tau~ a^3_{\rm F}
\left[2\Lambda_{\pm, {\rm F}}-\frac{6K}{a^2_{\rm F}}
 + m_g^2Y_\pm
 |X_\pm|\sqrt{\frac{K-\left(a_{\rm{F}}H_{\rm{F}}\right)^2}{K-\left(a_{\rm{F}}FX_\pm\right)^2}}\right]\,,\label{Sout}\\
S_{\rm wall} &\equiv \int {\rm d}^3x\sqrt{\Omega}
\int^{\tau_0(1+\delta)}_{\tau_0(1-\delta)}{\rm d}\tau~ a^3(\tau)
\left[2\big(V(\sigma)+\Lambda_{\pm}\big)-\frac{6K}{a^2(\tau)} +
m_g^2Y_\pm \sqrt{-(f^\prime)^2}\right]\,,\label{Sw}
\end{align}
with an infinitely small parameter $|\delta|\ll1$. In the following,
we use thin-wall approximation to evaluate
Eqs.~(\ref{Sin})--(\ref{Sw}).

%%%%%%%%%%%%%%%%%%%%
\subsection{Thin-wall approximation}\label{s:thinwallB}
%%%%%%%%%%%%%%%%%%%%
Recalling that the tunneling rate is expressed in terms of the
Euclidean action as shown in Eq.~(\ref{eq:rate}):
\begin{align}
    \Gamma\propto e^{-B}\,;
\quad
    B = S_{\rm E}[{g}_{\mu\nu,{\rm B}},{\phi}_{\rm B}]
-S_{\rm E}[{g}_{\mu\nu,{\rm F}},{\phi}_{\rm F}]\,, \label{eq:rateB}
\end{align}
where the exponential factor $B$ can be divided into three parts
with respect to the integration boundaries for $\tau$, in accordance
to the division of action $S_{E}$ in Eq.~(\ref{eq:actionCDL}):
\begin{eqnarray}\label{eq:B}
    B=B_{\rm inside}+B_{\rm outside}+B_{\rm wall}\,,
\end{eqnarray}
where
\begin{equation} \label{eq:Bres}
\left\{ \begin{aligned}
         B_{\rm inside} &\equiv S_{\rm{inside}}-S_{\rm F}|_{\tau<\tau_0}\,, \\
         B_{\rm outside} &\equiv S_{\rm{outside}}-S_{\rm F}|_{\tau>\tau_0}\,, \\
         B_{\rm wall} &\equiv S_{\rm{wall}}-S_{\rm F}|_{\tau=\tau_0}\,,
                          \end{aligned} \right.
                          \end{equation}
with $S_{\rm F}$ the corresponding Euclidean action of the false
vacuum. It immediately follows that $B_{\rm outside}=0$, since the
bouncing solution outside the bubble $\tau>\tau_0$ coincides with
that of false vacuum. So in the following, it is unnecessary to
evaluate the Euclidean action $S_{\rm outside}$.

Now we turn to evaluate $B_{\rm{inside}}$. From Eq.~(\ref{eq:eom}),
one obtains the following relationship:
\begin{align} \label{eq:adot}
a^{\prime} =\sqrt{K+\frac{a^2}{3}\left[\frac{\sigma^{\prime 2}}{2} -
V(\sigma) - \Lambda_{\pm}\right]}\,.
\end{align}
Inserting Eq.~(\ref{eq:adot}) into (\ref{Sin}) and rewrite the
integration as follows:
\begin{align} \label{eq:intein}
\int^{\tau_0(1-\delta)}_{0}{\rm
d}\tau~=\int^{a_0}_{0}\left(\diff{}{a}{\tau}\right)^{-1}{\rm d}a\,,
\end{align}
with $a_0\equiv a(\tau_0)$, then from Eq.~(\ref{eq:Bres}), the
exponential factor inside the bubble can be expressed as:
\begin{align}
B_{\rm{inside}} &=
2\pi^2K^{-\frac{3}{2}}\bigg\{\int^{a_0}_{0}\frac{a^3{\rm
d}a}{\sqrt{K-a^2\Lambda_{\pm, {\rm T}}/3}}~
 \left[2\Lambda_{\pm, {\rm T}}-\frac{6K}{a^2}
 + m_g^2Y_\pm |X_\pm|\sqrt{\frac{K-\left(aH_{\rm{T}}\right)^2}{K-\left(aFX_\pm\right)^2}}\right]\nonumber\\
 &-\int^{a_0}_{0}\frac{a^3{\rm
d}a}{\sqrt{K-a^2\Lambda_{\pm, {\rm F}}/3}}~
 \left[2\Lambda_{\pm, {\rm F}}-\frac{6K}{a^2}
 + m_g^2Y_\pm |X_\pm|\sqrt{\frac{K-\left(aH_{\rm{F}}\right)^2}{K-\left(aFX_\pm\right)^2}}\right]\bigg\}\nonumber\\
&=-12\pi^2K^{-\frac{3}{2}}\int^{a_0}_{0}a{\rm
d}a\left[\sqrt{K-(aH_{\rm T})^2}-\sqrt{K-(aH_{\rm
F})^2}\right]\,,\label{Bin}
\end{align}
where we have used $\int d^3x\sqrt{\Omega}=2\pi^2K^{-3/2}$ in the
first step and Eq.~(\ref{eq:hubblecdl}) in the last step. It should
be noted that though the term proportional to $m_g^2$ inside the
bubble (the last term in the first line of Eq.~(\ref{Bin}))
eliminates with the corresponding term outside (the last term in the
second line), nevertheless, the mass term contributes to the
effective cosmological constant $\Lambda_\pm$ as shown in
Eq.~(\ref{eq:friedCDL}), hence appears in the corresponding Hubble
parameters $H_{\rm T}$ and $H_{\rm F}$ by Eq.~(\ref{eq:hubblecdl}).

In order to evaluate the exponential factor on the wall $B_{\rm
wall}$, we use the thin-wall approximation~\cite{Coleman:1980},
\begin{eqnarray}\label{eq:TW}
    \mathcal{H}\sigma^\prime\ll1\,,
\end{eqnarray}
hence, Eq.~(\ref{eq:eom1}) can be easily solved as:
\begin{eqnarray}\label{eq:derisig}
    \sigma^\prime\simeq\sqrt{2\big[V(\sigma)-V(\sigma_{\rm
    T})\big]}\,.
\end{eqnarray}

Using the relationship
\begin{equation}
{\rm d}\tau=\left(\frac{{\rm d}\sigma}{{\rm d}\tau}\right)^{-1}{\rm
d}\sigma\,,
\end{equation}
inserting Eq.~(\ref{eq:derisig}) into the relationship above, the
exponential factor on the wall can be evaluated as follows:
\begin{align}
S_{\rm wall} &\simeq 2\pi^2K^{-\frac{3}{2}}~ \int_{\sigma_{\rm
T}}^{\sigma_{\rm F}}\frac{a^3_0{\rm
d}\sigma}{\sqrt{2\big[V(\sigma)-V(\sigma_{\rm
    T})\big]}}
\left[2\big(V(\sigma)+\Lambda_{\pm}\big)-\frac{6K}{a^2_0} +
m_g^2Y_\pm \sqrt{-
(f^\prime)^2}\bigg|_{\tau<\tau_0}\right]\,,\label{Swall}
\end{align}
while the Euclidean action of false vacuum on the wall can be
written as:
\begin{align}
S_{\rm F}|_{\tau=\tau_0} &\simeq 2\pi^2K^{-\frac{3}{2}}~
\int_{\sigma_{\rm T}}^{\sigma_{\rm F}}\frac{a^3_0{\rm
d}\sigma}{\sqrt{2\big[V(\sigma)-V(\sigma_{\rm
    T})\big]}}
\left[2\big(V(\sigma_{\rm{F}})+\Lambda_{\pm}\big)-\frac{6K}{a^2_0} +
m_g^2Y_\pm \sqrt{-
(f^\prime)^2}\bigg|_{\tau>\tau_0}\right]\,,\label{SwallF}
\end{align}
then inserting the above two equations into Eq.~(\ref{eq:Bres}),
under the assumption that
\begin{align}\label{TWcondition}
V(\sigma_{\rm HM})-V(\sigma_{\rm T})\gg V(\sigma_{\rm
F})-V(\sigma_{\rm T})\equiv\epsilon M_{\rm Pl}^4, \qquad
\epsilon\ll1,
\end{align}
one obtains the tunneling rate factor on the wall:
\begin{align}\label{Bwall}
B_{\rm wall} = 2\pi^2a^3_0K^{-\frac{3}{2}}~ \int_{\sigma_{\rm
T}}^{\sigma_{\rm F}}{\rm d}\sigma\sqrt{2\big[V(\sigma)-V(\sigma_{\rm
    T})]} + \mathcal{O}\left(\epsilon\right)\,.
\end{align}

Hence, combining Eqs.~(\ref{Bin}) and (\ref{Bwall}), the whole
tunneling rate factor can be expressed as
\begin{align}\label{Ball}
B \simeq-2\pi^2K^{-\frac{3}{2}}\left\{6\int^{a_0}_{0}a{\rm
d}a\left[\sqrt{K-(aH_{\rm T})^2}-\sqrt{K-(aH_{\rm F})^2}\right]-
a^3_0\Sigma\right\}\,,
\end{align}
where the tension $\Sigma$ is defined as follows:
\begin{align}\label{tension}
\Sigma\equiv\int_{\sigma_{\rm T}}^{\sigma_{\rm F}}{\rm
d}\sigma\sqrt{2\big[V(\sigma)-V(\sigma_{\rm
    T})]}\,,
\end{align}
and $a_0$ is determined by demanding that $B$ is stationary:
\begin{align}\label{eq:Bcri}
   \frac{{\rm d}B}{{\rm d}a_0}=0\quad\Longrightarrow\quad \frac{a_0}{2}\Sigma=\sqrt{K-(a_0H_{\rm T})^2}-\sqrt{K-(a_0H_{\rm F})^2}\,.
\end{align}

Thus, comparing Eqs.~(\ref{Ball}) and (\ref{eq:Bcri}) to the case in
GR~\cite{Coleman:1980}, we find that in thin-wall limit, provided
with the same value of $a_0$, $H_{\rm T}$ and $H_{\rm F}$,
respectively, the tunneling rate for the CDL instantons in nonlinear
massive gravity is the same as the one in GR. However, investigation
of HM instantons in dRGT massive gravity theory shows contributions
to tunneling rate coming from the graviton mass~\cite{ZSS:2013}.
Hence, in the next section, we take another limit of solutions
--- ``thick wall'' approximation --- to investigate the CDL solution.

%%%%%%%%%%%%%%
\section{CDL solution as perturbations around Hawking-Moss
solution}\label{s:CDLperturb}
%%%%%%%%%%%%%%

%%%%%%%%%%%%%%
\subsection{A brief summary of Hawking-Moss instanton in nonlinear massive
gravity}\label{HMcase}
%%%%%%%%%%%%%%

The Hawking-Moss (HM) instanton in nonlinear massive gravity has
been discussed in details in Ref.~\cite{ZSS:2013}. In this
subsection, we make a brief review on the results.

A HM solution can be found by setting the tunneling field to the
local maximum value, $\sigma(\xi) = \sigma_{\rm HM}$, as illustrated
in Fig.~\ref{fig:potential}.
 Then the equation of motion (\ref{eq:eom1}) is trivially satisfied
and the Euclidean Friedmann equation (\ref{eq:eom}) reduces to
    \begin{align}\label{eq:eom2}
        3\left(\frac{a^\prime}{a}\right)^2 - \frac{3K}{a^2}
 = - V(\sigma_{\rm HM}) - \Lambda_{\pm} \equiv -\Lambda_{\pm, \rm{eff}}\,.
     \end{align}
Setting the boundary condition $a_{\rm HM}(H_{\rm
HM}\tau=\pm\pi/2)=0$ and assuming $\Lambda_{\pm, \rm{eff}}>0$, the
HM solution is obtained as
\begin{align}\label{eq:aHMsol}
   a_{\rm{HM}}(\tau)
= H_{\rm HM}^{-1}\sqrt{K}\cos\left(H_{\rm HM}\tau\right)\,,
\end{align}
where
\begin{eqnarray}\label{eq:hubblehm}
  H_{\rm HM} \equiv  \sqrt{\frac{\Lambda_{\pm,{\rm eff}}}{3}}\,,
\end{eqnarray}
then under the constraint equaiton (\ref{eq:II}), one obtains
\begin{align}\label{eq:bf}
        b_{\rm HM} = F^{-1}\sqrt{K}\cosh(Ff_{\rm HM}) = X_{\pm}a_{\rm{HM}}\,.
\end{align}

Taking derivative with respect to $\tau$ on both sides of
Eq.~(\ref{eq:bf}), it immediately follows that:

\begin{align}\label{eq:sqfHM2der}
\left( f_{\rm HM}^\prime\right)^2 = \frac{X_{\pm}^2\sin^2(H_{\rm
HM}\tau)}{\alpha_{\rm HM}^2\cos^2(H_{\rm HM}\tau)-1}\,.
\end{align}
where the parameter $\alpha$ is defined as
\begin{eqnarray}\label{eq:reparameter}
        \alpha_{\rm HM} \equiv  X_\pm\frac{F}{H_{\rm HM}}\,.
\end{eqnarray}

Provided that $X_\pm>0$, then the parameter $\alpha_{\rm HM}>0$.
Moreover, we note that from Eq.~(\ref{eq:sqfHM2der}), it is clear
that at range $H_{\rm HM}\tau\in(-\pi/2, \pi/2)$, singularities will
appear unless $\alpha_{\rm HM}\leq1$. Hence, for consistency of the
theory, we derive the constraint:
\begin{eqnarray}\label{eq:alpharange}
        0<\alpha_{\rm HM}\leq1\,.
\end{eqnarray}

Inserting Eq.~(\ref{eq:eom2}) into the Euclidian action
Eqs.~(\ref{eq:euclidean}) and (\ref{eq:ematter}), and using
$N^{-1}\dot f={\rm d}f/{\rm d}\tau$, the total action can be
expressed as
    \begin{align}\label{eq:actionHM}
        S_{\rm{E, HM}} &= \int {\rm d}^3x\sqrt{\Omega}
\int^{\pi/2H_{\rm HM}}_{-\pi/2H_{\rm HM}}{\rm d}\tau~ a_{\rm HM}^3
\left(2\Lambda_{\pm,\rm{eff}}-\frac{6K}{a_{\rm HM}^2}
 + m_g^2Y_\pm \sqrt{-\left(f_{\rm HM}^\prime\right)^2}\right)\nonumber\\
 &= -\frac{8\pi^2}{H_{\rm HM}^2}\left[1 - \frac{Y_{\pm}X_\pm}{6}
\left(\frac{m_g}{H_{\rm HM}}\right)^2A(\alpha_{\rm HM})\right]\,,
    \end{align}
where the function $A(\alpha)$ is defined as follows:
\begin{align}\label{eq:A}
\quad A(\alpha)\equiv
\frac{2-\sqrt{1-\alpha^2}(2+\alpha^2)}{\alpha^4}\,.
\end{align}

It is clear from Eq.~(\ref{eq:actionHM}) that provided with the same
HM Hubble parameter, the second term proportional to $Y_\pm$ is the
correction term arising from the non-vanishing graviton mass. Hence,
compared with the corresponding Hawking-Moss tunneling rate in GR:
$B^{\rm (GR)}_{\rm HM}\equiv 8\pi^2(-H_{\rm HM}^{-2}+H_{\rm
F}^{-2})$, one obtains the correction term arising from the mass of
graviton:
\begin{eqnarray}\label{eq:deltaB}
\Delta B_{\rm HM}\equiv B^{\rm (MG)}_{\rm HM}-B^{\rm (GR)}_{\rm
HM}=\frac{4\pi^2m_g^2}{3}Y_{\pm}X_{\pm}\left[\frac{A(\alpha_{\rm
HM})}{H_{\rm HM}^4} -\frac{A(\alpha_{\rm F})}{H_{\rm F}^4}\right]\,.
\end{eqnarray}

Since the function $A(\alpha)$ is both positive and monotonically
increasing for $0<\alpha\leq1$, we have $A(\alpha_{\rm
F})>A(\alpha_{\rm HM})>0$. Together with $H_{\rm HM}>H_{\rm F}$, we
find that the sign of $\Delta B_{\rm HM}$ is determined by that of
$Y_\pm$, i.e. for $Y_{\pm}>0$ ($<0$), the correction $\Delta B<0$
($>0$), which implies that the tunneling rate is enhanced
(suppressed) compared to the case of GR.

%%%%%%%%%%%%%%
\subsection{Perturbations around Hawking-Moss solution}\label{perturb}
%%%%%%%%%%%%%%

The CDL solution can be also investigated as perturbations around
the HM solution~\cite{Tanaka:1992}. Firstly, we expand the potential
$V(\sigma)$ around $\sigma=\sigma_{\rm HM}$ as follows,
    \begin{eqnarray}\label{eq:potential}
V(\sigma)=V(\sigma_{\rm HM})-\frac{M^2}{2}(\sigma-\sigma_{\rm
HM})^2+\frac{m}{3}(\sigma-\sigma_{\rm
HM})^3+\frac{\nu}{4}(\sigma-\sigma_{\rm HM})^4 + \cdots\,,
    \end{eqnarray}
where we have introduced $M^2 \equiv -{\rm d}^2V(\sigma)/{\rm
d}\sigma^2|_{\sigma=\sigma_{\rm HM}}$. Near the HM limit where
$M^2\equiv4H_{\rm HM}^2(1+\chi^2)$ with $\chi^2\ll1$, the regular
solutions are perturbatively found to be
\begin{eqnarray}\label{eq:perturba}
a(\tau)&=&\tilde{H}_{\rm HM}^{-1}\cos\left(\tilde{H}_{\rm
HM}\tau\right)\left[1+\frac{\varepsilon_M^2H_{\rm
HM}^2}{8}\cos^2\left(\tilde{H}_{\rm HM}\tau\right) \right]
+ \mathcal{O}(\varepsilon_M^3)\\
&\equiv& a_0(\tau)+\delta a(\tau)\,,\nonumber\\
\sigma(\tau) &=&\sigma_{\rm HM}+\varepsilon_M H_{\rm
HM}\sin\left(\tilde{H}_{\rm
HM}\tau\right)+\frac{\varepsilon_M^2m}{12}\left[1-2\sin^2\left(\tilde{H}_{\rm
HM}\tau\right)\right]\nonumber\\
&-&\!\!\varepsilon_M^3H_{\rm HM}\sin\left(\tilde{H}_{\rm
HM}\tau\right)\!\!\left[\frac{3H_{\rm
HM}^2-4\mu}{56}\cos^2\left(\tilde{H}_{\rm
HM}\tau\right)-\frac{m^2}{36H_{\rm HM}^2}\sin^2\left(\tilde{H}_{\rm
HM}\tau\right)\right]\!+\!\mathcal{O}(\varepsilon_M^4),\label{eq:perturbscalar}
\end{eqnarray}
where $\tilde{H}_{\rm HM} \equiv H_{\rm HM}(1+H_{\rm
HM}^2\varepsilon_M^2/24)$, $\varepsilon_M^2\equiv84\chi^2/(16H_{\rm
HM}^2+9\mu)$ and $\mu\equiv\nu+m^2/18H_{\rm HM}^2$, while for
simplicity of symbols, we define the background value
$a_0(\tau)\equiv\tilde{H}_{\rm HM}^{-1}\cos\left(\tilde{H}_{\rm
HM}\tau\right)$ and the perturbations around it as $\delta a(\tau)$.
As in the case of GR, perturbations of Euclidean Hilbert-Einstein
action vanish up to order $\varepsilon_M^4$. Hence, up to order
$\varepsilon_M^2$, it is sufficient to evaluate the mass part of the
action:
\begin{eqnarray}\label{dsformal}
\delta S &=& - m_g^2\delta\left\{\int\mr{d}^4x_E~\sqrt{\Omega}
 \left(\mathcal{L}_{2E}+\alpha_3\mathcal{L}_{3E}+\alpha_4\mathcal{L}_{4E}\right)\right\}\nonumber\\
 &=&
-2\pi^2m_g^2~\delta\left\{\int_{-\pi/2\tilde{H}_{\rm
HM}}^{\pi/2\tilde{H}_{\rm HM}} {\rm d}\tau\sqrt{-
(f^\prime)^2}(a-b)\big[-3a^2-3\alpha_3
a(a-b)-\alpha_4(a-b)^2\big]\right\}\nonumber\\
&=& 4\pi^2m_g^2Y_\pm\int_0^{\pi/2\tilde{H}_{\rm HM}} {\rm
d}\tau~a_{\rm 0}^2\left[~3\sqrt{-(f_0^\prime)^2}~\delta a+a_{\rm
0}\delta\sqrt{- (f^\prime)^2}~\right]\,,
\end{eqnarray}
where in the last step, we have used the constraint equations
$b_{\rm 0}=X_\pm a_{\rm 0}$ and correspondingly, $\delta
b=X_\pm\delta a$. It is shown in Appendix~\ref{perturbHM} that by
using Eq.~(\ref{eq:perturba}), the second term $\delta\sqrt{-
(f^\prime)^2}$ can be expressed as follows:
\begin{eqnarray}\label{df1}
\delta\sqrt{- (f^\prime)^2}=\frac{\varepsilon_M^2X_\pm\tilde{H}_{\rm
HM}^2\sin\left(\tilde{H}_{\rm
HM}\tau\right)\cos^2\left(\tilde{H}_{\rm
HM}\tau\right)}{8\sqrt{1-\tilde{\alpha}^2\cos^2\left(\tilde{H}_{\rm
HM}\tau\right)}}\left[3+\frac{\tilde{\alpha}^2\cos^2\left(\tilde{H}_{\rm
HM}\tau\right)}{1-\tilde{\alpha}^2\cos^2\left(\tilde{H}_{\rm
HM}\tau\right)}\right]\,,
\end{eqnarray}
where $\tilde{\alpha}\equiv FX_\pm/\tilde{H}_{\rm HM}$. Thus,
inserting Eqs.~(\ref{eq:perturba})and (\ref{df1}) into
(\ref{dsformal}), one finally obtains the second order perturbation
in action as:
\begin{eqnarray}\label{s2}
\delta^{(2)}S =\frac{\pi^2m_g^2X_\pm Y_\pm{H}_{\rm
HM}^2\varepsilon_M^2}{2\tilde{H}_{\rm
HM}^4\sqrt{1-\tilde{\alpha}^2}}\,.
\end{eqnarray}

As can be seen above, since $X_\pm>0$, the sign of perturbation
depends on parameter $Y_\pm$ defined by Eq.~(\ref{eq:Y}). When
$Y_\pm>0$, $\delta^{(2)}S>0$ so that HM solution dominates, while if
$Y_\pm<0$, $\delta^{(2)}S<0$ so that the CDL solution dominates,
which is in sharp difference from the case of GR where the CDL
instanton always dominates over HM one, if it exists.

%%%%%%%%%%%%%%
\subsection{Beyond HM and thin-wall approximation}
%%%%%%%%%%%%%%
Generally speaking, it is difficult to analytically estimate the
tunneling rate beyond HM or thin-wall approximation. Nevertheless,
in this subsection, we present an estimation of the qualitative
behavior for a more general case.

Firstly, we rewrite the previous results in an uniform way. The mass
term can be written as follows,
%We recall that from Sec.~\ref{s:CDLthin}, in the thin-wall limit,
%the tunneling rate for CDL instantons is the same as that of GR, as
%shown in Eq.~(\ref{Ball}), while in Sec.~\ref{perturb}, by treating
%the CDL instantons as perturbations around HM solutions, correction
%term arises from the non-vanishing mass of the graviton as shown in
%Eq.~(\ref{s2}).
%%One may think that these two results contradict with
%%each other. However, in fact, the contradiction does not occur,
%%since the second approach (perturbations around HM solutions)
%%corresponds to a ``thick-wall'' limit, which is totally different
%%from the thin-wall limit.
%To interpret the results drawn from these two approaches, we
%concentrate on the non-vanishing mass term in the action
%(\ref{eq:euclidean}), which is the characteristic term of nonlinear
%massive gravity theory
\begin{align}\label{eq:mass}
        S^{{\rm mass}} &\equiv - m_g^2\int\mr{d}^4x_E~\sqrt{\Omega}
 \left(\mathcal{L}_{2E}+\alpha_3\mathcal{L}_{3E}+\alpha_4\mathcal{L}_{4E}\right)\nonumber\\
 &= 2\pi^2K^{-\frac{3}{2}}m_g^2Y_\pm
\int{\rm d}\tau~ a^3(\tau)
 \sqrt{- (f^\prime)^2}\,,
\end{align}
where in the second step we have inserted Eqs.~(\ref{eq:II}),
(\ref{eq:Y}) and $\int d^3x\sqrt{\Omega}=2\pi^2K^{-3/2}$. On the
other hand, %similar with the situation of HM solution
%(\ref{eq:aHMsol}), the Friedman equation~(\ref{eq:eom}) can be
%solved in the following way:
%\begin{align}\label{eq:friedsol}
%   a(\tau)
%= H^{-1}\sqrt{K}\cos\left(H\tau\right)\,,
%\end{align}
%with $H^2\equiv[-\sigma^{\prime 2}/2+V(\sigma)+\Lambda_\pm]/3$.
%Hence,
taking derivative of Eq.~(\ref{eq:bfa}) with respect to $\tau$, one
can express $f^\prime$ in terms of $a^\prime$ and $a$ as follows:
\begin{align}\label{eq:fdotsol}
-(f^\prime)^2 = \frac{X_\pm^2 (a^\prime)^2}{K-\left(FX_\pm
a\right)^2}\,.
\end{align}

Hence, inserting Eq.~(\ref{eq:fdotsol}) into (\ref{eq:mass}), one
can express $S^{{\rm mass}}$ as integration of $a$ instead of $\tau$
as follows
\begin{align}\label{eq:massa}
        S^{{\rm mass}}%&= 2\pi^2K^{-\frac{3}{2}}m_g^2X_\pm Y_\pm
%\int_{-\pi/2H}^{\pi/2H}{\rm d}\tau~ \frac{a^3|a^\prime|}{\sqrt{K-\left(FX_\pm a\right)^2}}\nonumber\\
&= 4\pi^2K^{-\frac{3}{2}}m_g^2X_\pm Y_\pm \int_{0}^{a_{\rm{max}}}~
\frac{a^3{\rm d}a}{\sqrt{K-\left(FX_\pm a\right)^2}}\nonumber\\
&= -\frac{4\pi^2K^{-\frac{3}{2}}m_g^2X_\pm Y_\pm}{3(FX_\pm)^4}
\left[\sqrt{K-\left(FX_\pm a\right)^2}\left(2K+\left(FX_\pm
a\right)^2\right)\right]_{0}^{a_{\rm max}}\,,
\end{align}
where $a_{\rm{max}}$ is the largest radius of the bubble in
Euclidean time. For convenience of discussion, we define $B^{\rm
mass}$ as follows
\begin{align}\label{eq:Bdefine}
B^{\rm mass}(a_{\rm max}) &\equiv S^{\rm mass}-S^{\rm mass}_{\rm
F}\propto \left[\sqrt{K-\left(FX_\pm
a\right)^2}\left(2K+\left(FX_\pm a\right)^2\right)\right]_{a_{\rm{F,
max}}}^{a_{\rm max}}\,,
\end{align}
which is the correction term arising from the graviton mass when one
calculates the tunneling rate by Eq.~(\ref{eq:rate}).

\begin{figure}
\includegraphics[height=5.5cm,keepaspectratio=true,angle=0]{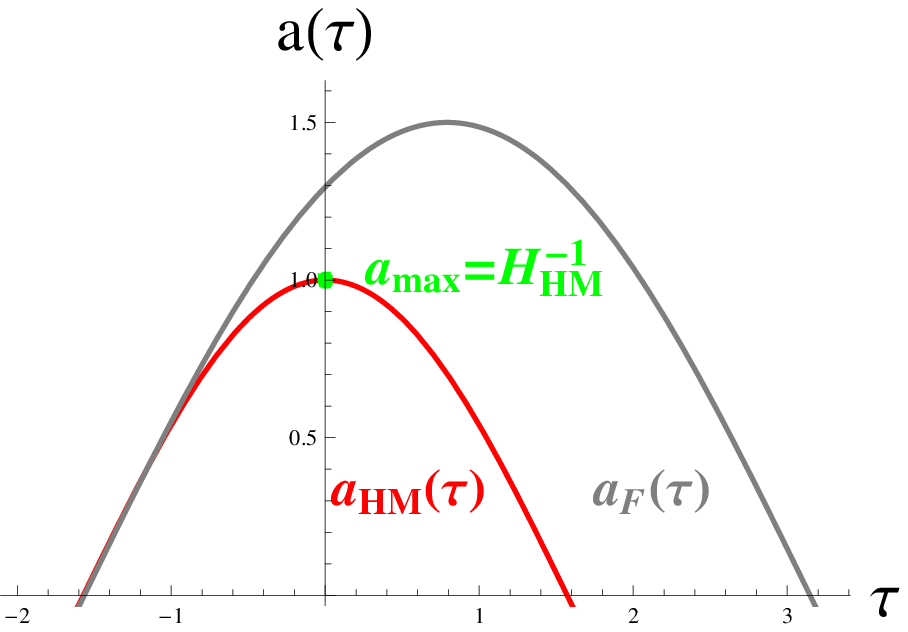}~~~~~~
\includegraphics[height=5.5cm,keepaspectratio=true,angle=0]{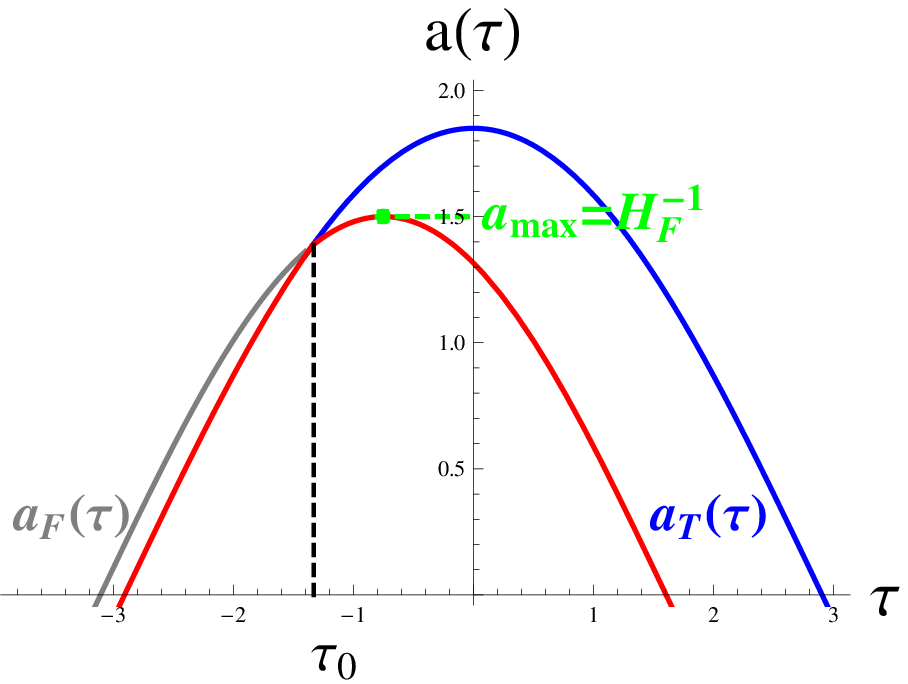}
\caption{Illustration of scale factor $a(\tau)$ in cases of
Hawking-Moss instanton (left panel) and Coleman-de Luccia instanton
(right panel). In both cases, the evolution of the scale factor is
labeled by red curves, while that of false vacuum $a_{\rm F}(\tau)$
and true vacuum $a_{\rm T}(\tau)$ are labeled by gray and blue
curves, respectively. From the left panel, it is obvious that the
maximum value of scale factor in HM case is different from that of
false vacuum, which leads to a non-vanishing contribution to the
tunneling rate as shown in Eq.~(\ref{eq:Binamaxhm}). However, from
the right panel, one finds that in the CDL case, the scale factor
firstly evolves along that of true vacuum until some point $\tau_0$,
then follows along that of false vacuum so that its maximum value
coincides with that of false vacuum. Hence, using
Eq.~(\ref{eq:Bdefine}), contribution from the graviton mass vanishes
in the CDL case.} \label{fig:amax}
\end{figure}

Now let us reconsider the conclusions drawn from thin-wall and HM
limit by evaluating $B^{\rm mass}$ in these two cases, respectively.
We recall that in the thin-wall limit, the maximum value of the
scale factor is equal to that of the false vacuum:
$a_{\rm{max}}=a_{\rm{F, max}}\equiv H_{\rm{F}}^{-1}$, as illustrated
in the right panel of Fig.~\ref{fig:amax}. Hence, using
Eqs.~(\ref{eq:Bres}) and (\ref{eq:massa}), one obtains a vanishing
correction term in the thin-wall limit:
\begin{align}\label{eq:Binamaxtw}
B^{\rm mass}_{\rm thin-wall}=B^{\rm mass}(a_{\rm max}=a_{\rm{F,
max}})=0\,,
\end{align}
which leads to the conclusion in Sec.~\ref{s:CDLthin} that the
non-vanishing mass of the graviton does not contribute to the CDL
tunneling rate in the thin-wall limit.

On the other hand, the HM limit corresponds to a ``thick-wall''
limit, where the maximum value of the scale factor is given by that
of the local maximum between true and false vacuum,
$a_{\rm{max}}=a_{\rm{HM, max}}\equiv
H_{\rm{HM}}^{-1}<H_{\rm{F}}^{-1}$ (see the left panel of
Fig.~\ref{fig:amax}), hence leads to a non-vanishing correction term
for tunneling rate:
\begin{align}\label{eq:Binamaxhm}
B^{\rm mass}_{\rm HM}=B^{\rm mass}(a_{\rm
max}=H_{\rm{HM}}^{-1})=\frac{4\pi^2m_g^2}{3}X_{\pm}Y_{\pm}\left[\frac{A(\alpha_{\rm
HM})}{H_{\rm HM}^4} -\frac{A(\alpha_{\rm F})}{H_{\rm F}^4}\right]\,,
\end{align}
where in the last step, we set $K=1$ while $\alpha$ and $A(\alpha)$
are defined in Eqs.~(\ref{eq:reparameter}) and (\ref{eq:A}),
respectively. It is obvious that Eq.~(\ref{eq:Binamaxhm}) coincides
with (\ref{eq:deltaB}) as expected.

Comparison of Eq.~(\ref{eq:Binamaxtw}) to (\ref{eq:Binamaxhm})
suggests the expectation that deviations from thin-wall and HM limit
may lead to contributions to the tunneling rate which change
monotonically in $a_{\rm max}$ in nonlinear massive gravity theory.
Hence, in the following, we consider deviations from thin-wall and
HM limit, respectively.

\subsubsection{Deviation from HM and thin-wall approximation}\label{deviTW}

The HM instanton corresponds to a ``thick-wall'' limit where the
curvature of local maximum of the potential is flat enough: ${\rm
d}^2V(\sigma_{\rm HM})/{\rm d}\sigma^2=4H_{\rm HM}^2$. Using the
perturbation approach in Sec.~\ref{perturb}, from
Eq.~(\ref{eq:perturba}), small deviation from HM instanton implies
that
\begin{eqnarray}\label{eq:amaxperturb}
a_{\rm max}\simeq H_{\rm perturb}^{-1}\equiv\tilde{H}_{\rm
HM}^{-1}\left(1+\frac{\varepsilon_M^2H_{\rm HM}^2}{8} \right)>a_{\rm
HM}\,.
\end{eqnarray}

Inserting Eq.~(\ref{eq:amaxperturb}) into (\ref{eq:massa}), the
correction term for the perturbational approach around HM limit is
evaluated as
\begin{align}\label{eq:Binamaxper}
B^{\rm mass}_{\rm perturb} &=B^{\rm mass}(a_{\rm max}=H_{\rm perturb}^{-1})\nonumber\\
&=-\frac{4\pi^2m_g^2X_\pm Y_\pm}{3(FX_\pm)^4}
\left[\sqrt{1-\tilde{\alpha}^2}\left(2+\tilde{\alpha}^2\right)-\sqrt{1-\alpha_{\rm
F}^2}\left(2+\alpha_{\rm F}^2\right)\right]+\frac{m_g^2\pi^2X_\pm
Y_\pm{H}_{\rm HM}^2\varepsilon_M^2}{2\tilde{H}_{\rm
HM}^4\sqrt{1-\tilde{\alpha}^2}}+\mathcal{O}(\varepsilon_M^4)\nonumber\\
&=-\frac{4\pi^2m_g^2X_\pm Y_\pm}{3(FX_\pm)^4}
\left[\sqrt{1-\alpha_{\rm HM}^2}\left(2+\alpha_{\rm
HM}^2\right)-\sqrt{1-\alpha_{\rm F}^2}\left(2+\alpha_{\rm
F}^2\right)\right]+\frac{m_g^2\pi^2X_\pm
Y_\pm\varepsilon_M^2}{3H_{\rm HM}^4\sqrt{1-\alpha_{\rm
HM}^2}}\nonumber\\
&\quad+\mathcal{O}(\varepsilon_M^4)\,.
\end{align}

Obviously, in the second step, the correction term of order
$\varepsilon_M^2$ coincides with Eq.~(\ref{s2}) as expected. Hence,
the result in Sec.~\ref{perturb} is recovered by applying
Eq.~(\ref{eq:massa}).
\\

\begin{figure}
\includegraphics[height=6.5cm,keepaspectratio=true,angle=0]{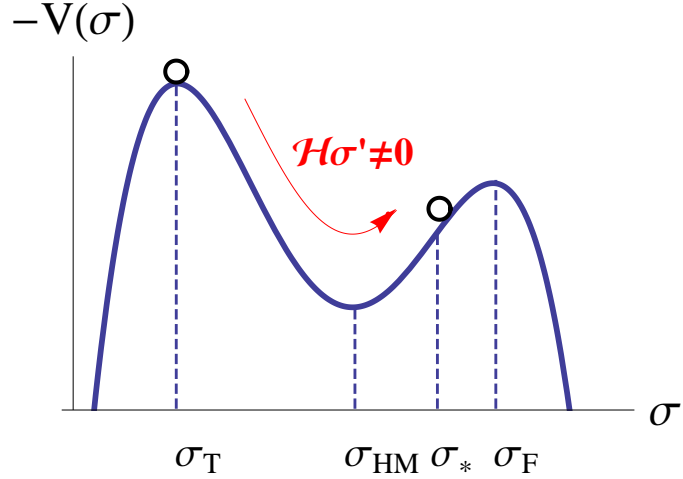}
\caption{Illustration of classical trajectory with $-V(\sigma)$.
When friction term $\mathcal{H}\sigma^\prime$ is taken into account,
scalar field starting around $\sigma_{\rm T}$ along the potential
cannot reach $\sigma_{\rm F}$ as in the thin-wall case. Instead, it
will stop at some point $\sigma_{\rm *}$ and correspondingly gives
the maximum radius $a_{\rm max}=a_{\rm *, max}<a_{\rm{F, max}}$ as
shown in Eq.~(\ref{eq:astar}), hence leads to correction of
tunneling rate arising from the non-vanishing graviton mass.}
 \label{fig:potential2}
\end{figure}

For an intuitive analysis of deviation from thin-wall approximation,
let us consider the classical trajectory of $-V(\sigma)$, where the
scalar field is driven from the true vacuum $\sigma_{\rm T}$ toward
the false vacuum $\sigma_{\rm F}$, as illustrated in
Fig.~\ref{fig:potential2}. In the thin-wall limit, the friction term
$3\mathcal{H}\sigma^\prime$ in Eq.~(\ref{eq:eom1}) is neglected so
that $\sigma$ can reach $\sigma_{\rm F}$ because of conservation of
total energy. However, small deviation from the thin-wall limit
implies a non-negligible friction term which causes loss of energy
so that the scalar field starting from $\sigma_{\rm T}$ can only
reach some point $\sigma_{\rm *}$ where $V(\sigma_{\rm
*})>V(\sigma_{\rm F})$. Correspondingly, to calculate the maximum
radius of the bubble, we have ${\rm d}a_{\rm *, max}/{\rm d}\tau=0$,
then inserting into Eq.~(\ref{eq:eom}), one can evaluate in the
following way:
\begin{align}\label{eq:astar}
a_{\rm *, max}=\sqrt{\frac{3}{V(\sigma_{\rm
*})+\Lambda_\pm}}<a_{\rm{F, max}}\equiv H_{\rm F}^{-1}\,.
\end{align}

Again inserting Eq.~(\ref{eq:astar}) into (\ref{eq:massa}), the
correction term can be evaluated as:
\begin{align}\label{eq:Binamaxstar}
B^{\rm mass}_{\rm *} &=B^{\rm mass}(a_{\rm max}=H_{\rm
*}^{-1})=\frac{4\pi^2m_g^2}{3}X_{\pm}Y_{\pm}\left[\frac{A(\alpha_{\rm
*})}{H_{\rm *}^4} -\frac{A(\alpha_{\rm F})}{H_{\rm F}^4}\right]\neq0\,.
\end{align}

Hence, small deviation from thin-wall limit makes $a_{\rm
max}<a_{\rm{F, max}}$, which furthermore leads to corrections to the
tunneling rate for the CDL instanton when compared to GR.

\subsubsection{Beyond HM and thin-wall approximation}

\begin{figure}
\includegraphics[height=6.5cm,keepaspectratio=true,angle=0]{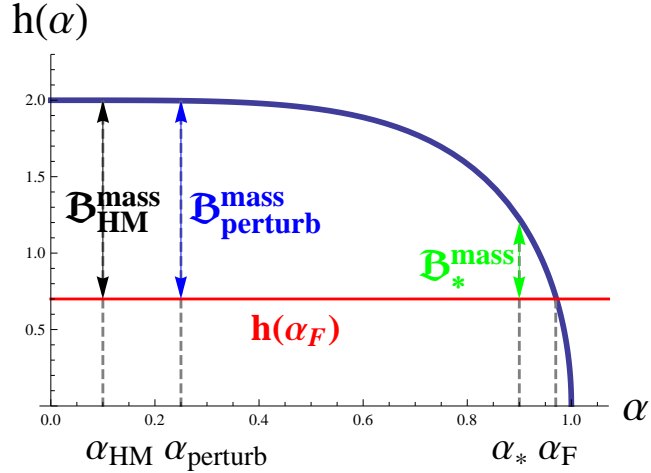}
\caption{Plot of function $h(\alpha)$ defined in Eq.~(\ref{eq:f}).
For a given model, $\alpha_{\rm HM}$ and $\alpha_{\rm F}$ are
definite, from which one obtains $h(\alpha_{\rm HM})$ and
$h(\alpha_{\rm F})$. Using Eq.~(\ref{eq:Bredefine}), the value of
normalized factor $\mathfrak{B}^{\rm mass}(a_{\rm max})$ is
illustrated by black, blue and green double arrow lines,
corresponding to the value in cases of HM (Sec.~\ref{HMcase}),
perturbations from HM (Sec.~\ref{perturb}) and deviation from
thin-wall limit (Sec.~\ref{deviTW}), respectively. It should be
noted that since $\alpha\in(0, 1]$, the value of $\mathfrak{B}^{\rm
mass}$ is of order unity: $\mathfrak{B}^{\rm mass}\in[0, 2)$.}
 \label{fig:Bmass}
\end{figure}

For a qualitative analysis, it is convenient to define a normalized
function $\mathfrak{B}^{\rm mass}$ in the following way:
\begin{align}\label{eq:Bredefine}
\mathfrak{B}^{\rm
mass}(a_{\rm{max}})&\equiv-\frac{3(FX_\pm)^4{B}^{\rm
mass}}{4\pi^2m_g^2X_\pm Y_\pm}=\left[\sqrt{1-\left(FX_\pm
a\right)^2}\left(2+\left(FX_\pm
a\right)^2\right)\right]_{H_{\rm{F}}^{-1}}^{a_{\rm{max}}}\nonumber\\
&\equiv h(\alpha_{\rm max})-h(\alpha_{\rm F})\geq0\,,
\end{align}
where we have defined the function $f$ and variable $\alpha$ as
follows:
\begin{eqnarray}\label{eq:f}
h(\alpha)\equiv \sqrt{1-\alpha^2}(2+\alpha^2),\qquad \alpha\equiv
FX_{\rm \pm}a\,,
\end{eqnarray}
which is plotted in Fig.~\ref{fig:Bmass}. It should be noted that
$\alpha_{\rm F}\equiv FX_{\rm \pm}/H_{\rm F}$ coincides with the one
in Ref.~\cite{ZSS:2013} and leads to a constraint on the height of
potential at false vacuum:
\begin{align}\label{eq:paraahphaf}
\alpha_{\rm F}\leq1\quad\Longrightarrow\quad V(\sigma_{\rm
F})\geq3F^2X_{\pm}^2-\Lambda_{\pm}\,.
\end{align}

Inserting Eq.~(\ref{eq:Bredefine}) into (\ref{eq:Bdefine}), the
correction to the tunneling rate in dRGT massive gravity can be
expressed as:
\begin{align}\label{eq:Bredefine1}
\Delta\Gamma\equiv\frac{\Gamma_{\rm MG}}{\Gamma_{\rm
GR}}\simeq\exp(-B^{\rm mass})=\exp\left(\frac{4\pi^2m_g^2
Y_\pm\mathfrak{B}^{\rm mass}}{3F^4X_\pm^3}\right)\,.
\end{align}

Provided with definite parameters $m_g^2$, $\alpha_3$ and
$\alpha_4$, it is convenient to consider the behavior of
$\mathfrak{B}^{\rm mass}$ with respect to $a_{\rm max}$, as plotted
in Fig.~\ref{fig:Bmass}. The values of normalized factor
$\mathfrak{B}^{\rm mass}(a_{\rm max})$ defined in
Eq.~(\ref{eq:Bredefine}) are illustrated by black, blue and green
double arrow lines, corresponding to the value in cases of HM
(Sec.~\ref{HMcase}), perturbations from HM (Sec.~\ref{perturb}) and
deviation from thin-wall limit (Sec.~\ref{deviTW}), respectively. In
the thin-wall limit $a_{\rm max}=H_{\rm F}^{-1}$ (i.e.
$\alpha=\alpha_{\rm F}$), hence the graviton mass has no
contribution to the CDL tunneling rate as shown in
Eq.~(\ref{eq:Binamaxtw}). Provided with $Y_{\pm}>0$, under
deviations from thin-wall limit, $a_{\rm max}=H_{\rm
*}^{-1}<H_{\rm F}^{-1}$ (i.e. $\alpha=\alpha_{\rm *}$). Hence, a non-vanishing
contribution to the CDL tunneling rate arises (as shown in the green
double arrow line) and increases gradually, as evaluated in
Eq.~(\ref{eq:Binamaxstar}).

On the other hand, the most probable point is the HM limit which can
be interpreted as ``thick-wall'' limit and correspondingly gives the
largest tunneling rate as expected (black double arrow line).
Perturbations around this limit give larger maximum radius $a_{\rm
max}=H_{\rm perturb}^{-1}>H_{\rm HM}^{-1}$ (i.e. $\alpha=\alpha_{\rm
perturb}$) as shown in Eq.~(\ref{eq:amaxperturb}), so the tunneling
rate decreases monotonically (blue double arrow line). Thus,
Eq.~(\ref{eq:Bredefine}) implies monotonic behavior of function
$\mathfrak{B}^{\rm mass}$ when $a_{\rm max}$ changes from $H_{\rm
HM}^{-1}$ to $H_{\rm F}^{-1}$. Correspondingly, a monotonic behavior
of the CDL tunneling rate with respect to different $a_{\rm max}$ is
expected.

It should be noted that if $Y_\pm<0$, the above conclusion holds
inversely, while $Y_\pm=0$ implies vanishing contribution to the
tunneling rate. Since $0<\alpha\leq1$, the value of
$\mathfrak{B}^{\rm mass}$ is of order unity: $0\leq\mathfrak{B}^{\rm
mass}<2$.

Moreover, when $F\longrightarrow0$, the fiducial metric becomes
Minkowskian. In this case, Eq.~(\ref{eq:Bredefine1}) reduces to the
following form
\begin{align}\label{eq:BredefineMinkowski}
\Delta\Gamma=\exp\left[\pi^2X_\pm Y_\pm m_g^2 \left(H_{\rm
F}^{-4}-a_{\rm max}^4\right)\right]\,.
\end{align}

Similarly as the behavior in de Sitter fiducial metric case as shown
in Eq.~(\ref{eq:Bredefine1}), when $Y_\pm>0$, contribution to the
CDL tunneling rate arising from the graviton mass appears when one
go beyond ``thin-wall'' approximation, and increases monotonically
until its maximum value at HM point. If $Y_\pm<0$, the conclusion
holds inversely.

%%%%%%%%%%%%%%
\section{CDL v.s. HM process}\label{s:HMvsCDL}
%%%%%%%%%%%%%%

In the previous section, it is found that when compared to the
situation in GR, correction to the CDL tunneling rate will appear
because of the non-vanishing graviton mass, and its value will
change monotonically with respect of $a_{\rm max}$ until HM
solution, i.e. from ``thin-wall'' to ``thick-wall'' case.
Correspondingly, the physical picture of the analysis is that the
shape of the potential changes gradually: the rate of its typical
height of the local maximum to its width decreases monotonically.

%Furthermore, it is interesting to consider the case that provided
%with the same shape of potential, which process will dominate the
%tunneling process. For definiteness, we assume that the potential
%satisfies the ``thin-wall'' approximation so that the maximum value
%$a_{\rm max}$ coincides with that of false vacuum $H_{\rm F}^{-1}$,
%hence correction of Euclidean action due to the mass term eliminates
%with that of false vacuum, as shown in Eq.~(\ref{eq:Binamaxtw}).
%Deviation from ``thin-wall'' approximation shows appearance of
%correction of tunneling rate, the absolute value of which
%monotonically increases as the width of the wall grows thicker until
%HM case, as illustrated in Fig.~\ref{fig:Bmass}.

On the other hand, it is interesting to consider another prospect:
provided with the same shape of potential which satisfies the
``thin-wall'' condition Eq.~(\ref{TWcondition}), whether CDL
instanton will dominate over HM one or inversely, which may imply
sharp difference from GR where the CDL one always dominates if it
exists. The comparison of the probability of the CDL process to that
of HM is expressed as follows (for details of deduction see
Appendix~\ref{cdltoHM}):
\begin{align}\label{Bcdlhm}
\ln\left(\frac{P_{\rm CDL}}{P_{\rm
HM}}\right)\approx4\pi^2\left[\frac{16M_{\rm
{Pl}}^6}{\Sigma^2}-\frac{m_g^2M_{\rm Pl}^2X_\pm Y_\pm }{3}\left(
\frac{A(\alpha_{\rm F})}{H_{\rm F}^4}-\frac{A(\alpha_{\rm
HM})}{H_{\rm HM}^4}\right)\right]\,,
\end{align}
assumming $\Lambda_{\pm, \rm T}=0$ and $\Lambda_{\pm, \rm
F}=\epsilon M_{\rm Pl}^4$ with $\epsilon\ll1$, where $A(\alpha)$ is
defined in Eq.~(\ref{eq:A}). In the context of GR where $m_g=0$,
under the thin-wall approximation, the probability of the CDL
instanton always dominates over the HM one. However, in dRGT massive
gravity theory, there appears a term which is proportional to the
mass of graviton, hence gives rise to the possibility that HM
process may dominate over the CDL one when $Y_{\pm}>0$. It should be
noted that, even when $F \to 0$, where the fiducial metric reduces
to Minkowskian one, the function $A(\alpha)$ is finite,
$\lim_{\alpha \to 0} A(\alpha)=3/4$, and then the ratio
(\ref{Bcdlhm}) is non-singular as has been stated in
Ref.~\cite{ZSS:2013}.

To find such a case, we note that within the range $\alpha\in(0,
1]$, the function $A(\alpha)$ of order unity. Moreover, the
thin-wall approximation implies that $H_{\rm F}^{-4}=9/(\epsilon^2
M_{\rm Pl}^4)\gg H_{\rm HM}^{-4}$. Hence, provided that the
parameters $\alpha_3\sim\alpha_4\sim\mathcal{O}(1)$, one finds the
condition on the value of graviton mass for HM process dominance:
\begin{align}\label{mgconstr}
m_g>\mathcal{O}\left(M_{\rm {Pl}}^2H_{\rm
F}^2\Sigma^{-1}\right)\sim\mathcal{O}\left(a_0^{-1}\right)\,,
\end{align}
where $a_0$ is the radius of bubble defined in Eq.~(\ref{eq:Bcri})
while Eq.~(\ref{eq:Bcriapp}) has been used in the last step.

%It should be noted that when $F\longrightarrow0$, the fiducial
%metric reduces to Minkowskian one, which is non-singular as has been
%stated in Ref.~\cite{ZSS:2013}. In fact, in this case,
%Eq.~(\ref{Bcdlhm}) reduces to the following form
%\begin{align}\label{BcdlhmMinkowski}
%\lim_{F\rightarrow0}\left[\ln\left(\frac{P_{\rm CDL}}{P_{\rm
%HM}}\right)\right]&\approx4\pi^2\left[\frac{16}{\Sigma^2}+\frac{1}{4}m_g^2X_\pm
%Y_\pm\left(H_{\rm HM}^{-4}-H_{\rm
%F}^{-4}\right)\right]\nonumber\\
%&\approx4\pi^2\left(\frac{16}{\Sigma^2}-\frac{m_g^2X_\pm
%Y_\pm}{4H_{\rm F}^4}\right)\,,
%\end{align}
%which implies the same condition as Eq.~(\ref{mgconstr}) for HM
%domination.

\begin{figure}
\includegraphics[height=6.5cm,keepaspectratio=true,angle=0]{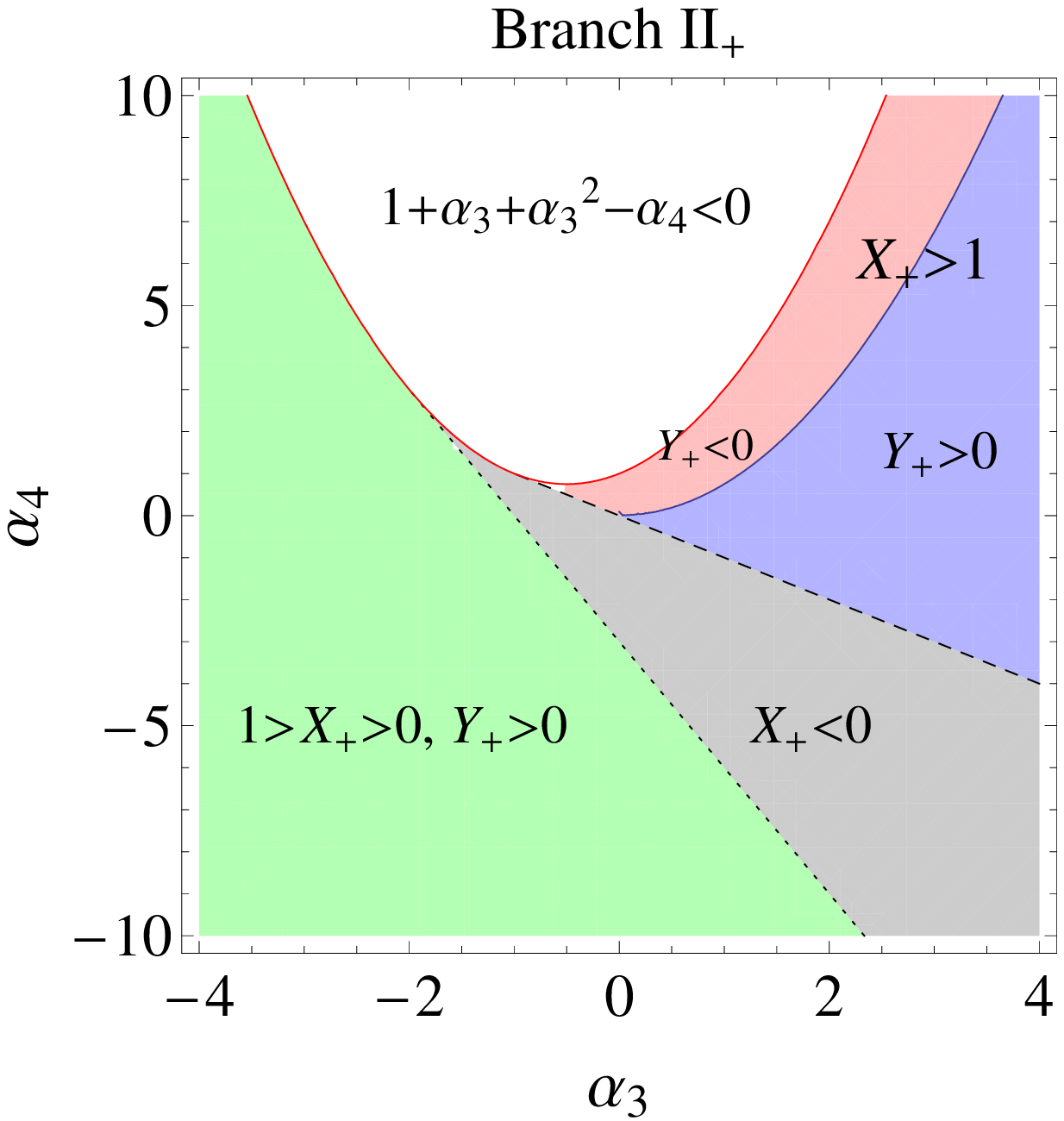}
\includegraphics[height=6.5cm,keepaspectratio=true,angle=0]{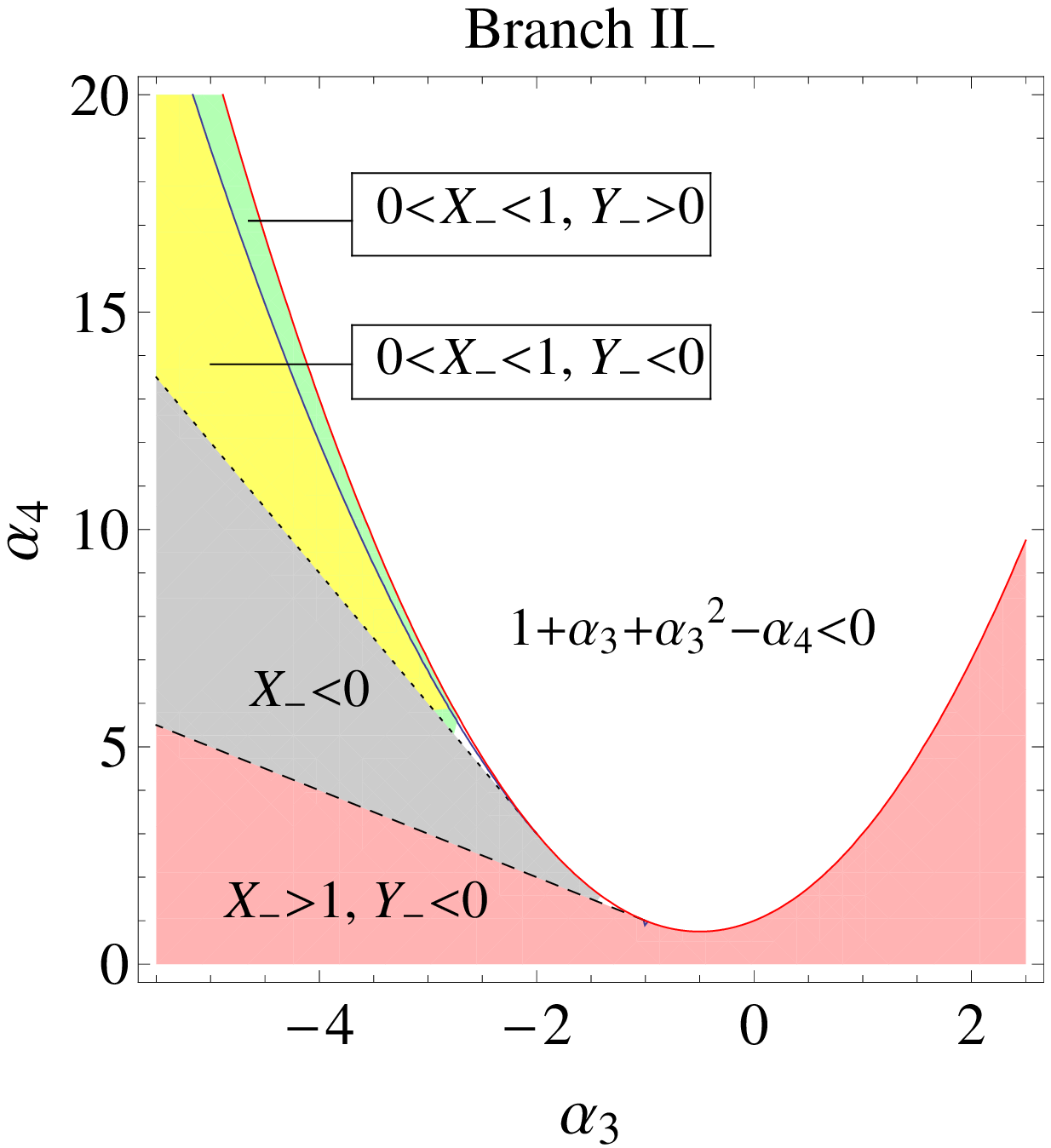}
\caption{The sign of $Y_\pm$ is shown for Branch II$_{+}$ (left
panel) and Branch II$_{-}$ (right panel) solutions, respectively.
The white region corresponds to $1+\alpha_3+\alpha_3^2-\alpha_4<0$
while in the gray one $X_\pm<0$, both of which should be excluded
since the cosmological solution does not exist in these
regions~\cite{Gumrukcuoglu:2011open, Gumrukcuoglu:2011perturb}. In
pink region, $X_\pm>1$ and $Y_\pm<0$, while green region corresponds
to $1>X_\pm>0$ with $Y_\pm>0$. In blue region, $X_+>1$ and $Y_+>0$,
while yellow region corresponds to $1>X_->0$ and $Y_->0$. Along the
solid lines (defining the boundary between blue and pink regions for
II$_{+}$, while the boundary between pink and yellow regions for
II$_{-}$), $Y_\pm=0$ so that it reduces to the case in GR. The
dotted lines denote $X_\pm=0$ where the solution ceases to exist.
Along the dashed lines, $X_\pm$ diverges and thus defines another
boundary of the solution space.} \label{fig:Xbranches}
\end{figure}

In order to see the possibilities for $Y_\pm>0$, in
Fig.~\ref{fig:Xbranches}, we show the sign of $Y_\pm$ in the
parameter space $(\alpha_3,\alpha_4)$ for Branch II$_{+}$ (left
panel) and Branch II$_{-}$ (right panel) solutions, respectively.
The value of parameter $X_\pm$ is further constrained by the mass of
tensor mode for self-accelerating solutions
~\cite{Gumrukcuoglu:2011perturb}
\begin{equation} \label{eq:tensormass}
M_{\rm {GW}}^2=\left\{ \begin{aligned}
          &  m_g^2X_+(1-X_+)\sqrt{1+\alpha_3+\alpha_3^2-\alpha_4}~,\quad {\rm for}~~ X_+\\
          &  m_g^2X_-(X_--1)\sqrt{1+\alpha_3+\alpha_3^2-\alpha_4}~,\quad {\rm for}~~
          X_-
                                    \end{aligned} \right.
                          \end{equation}
which implies the constraints on parameter $X_\pm$ to avoid the
tachyonic instability: in Branch II$_{+}$, $0<X_+<1$ and in Branch
II$_{-}$, $X_->1$. Hence, as shown in Fig.~\ref{fig:Xbranches},
there exists a region for $Y_+>0$ within the constraints (the green
region in Branch II$_{+}$) while in Branch II$_{-}$, under the
constraint $X_->1$, the parameter $Y_-$ is always negative.

Hence, in Branch II$_{+}$, there exists the case where HM process
would dominate over the CDL one, provided with Eq.~(\ref{mgconstr}).
However, in Branch II$_{-}$, the CDL process always dominates.
Moreover, it should be noted that in the limit where
$\alpha_3=\alpha_4=0$, $X_+$ diverges so that only Branch II$_{-}$
solution exists~\cite{ZSS:2013}.

Thus, under the thin-wall approximation where the potential is very
``sharp'' at its local maximum, when the value of graviton mass is
large enough so that Eq.~(\ref{mgconstr}) is satisfied, the HM
process may dominate over the CDL one, which is very different from
the case of GR. So we conclude that in the context of dRGT massive
gravity theory, not only the shape of the potential, but also the
values of parameters $\alpha_3$, $\alpha_4$ and $m_g$ will influence
the tunneling process.

\section{Conclusions}\label{conclusion}
%%%%%%%%%%%%%%

Towards the understanding of stability of vacuum in the landscape of
vacua in dRGT massive gravity, in this paper, we investigated the
Coleman-DeLuccia (CDL) solution, under the assumption that a
tunneling field minimally couples to gravity. For comparison with
Hawking-Moss (HM) instanton~\cite{ZSS:2013}, we choose
Branch-II$_{\pm}$ for analysis and evaluate the corresponding
tunneling rate of the CDL instanton. Firstly, we used the
``thin-wall'' approximation~\cite{Coleman:1980} and found that the
non-vanishing graviton mass term does not contribute to the
tunneling rate.

To compare this result with the HM case, where the non-vanishing
correction arises~\cite{ZSS:2013}, we derived the CDL solution as
perturbations around the HM case, which corresponds to a
``thick-wall'' approximation (or equivalently, the potential is very
``flat'' around its local maximum). In this approach, we found
non-vanishing second-order perturbation to the tunneling rate due to
the non-vanishing graviton mass, as shown in Eq.~(\ref{s2}), which
implies correction to the tunneling rate for the CDL instanton.
Moreover, it is found that in this approach, when the parameter
$Y_\pm>0$ (defined in Eq.~(\ref{eq:Y})), HM process will dominate
over the CDL one even when the CDL solutions exist.

%Provided with the HM solution obtained in~\cite{ZSS:2013}, as a
%different approach to the evaluation of tunneling rate, we derived
%CDL solution as perturbations around HM case, which corresponds to a
%``thick-wall'' approximation (or equivalently, the potential is very
%``flat'' around its local maximum). In this approach, we found
%non-vanishing second-order perturbation to the tunneling rate due to
%the non-vanishing mass of graviton, as shown in Eq.~(\ref{s2}),
%which implies correction to the tunneling rate for CDL instanton.
%Moreover, it is found that in this approach, when the parameter
%$Y_\pm>0$ (defined in Eq.~(\ref{eq:Y})), HM process will dominate
%over CDL one; when $Y_\pm<0$, CDL process will dominate over HM one.

In order to go beyond the thin-wall and thick-wall approximations,
we rewrite the corrections to the tunneling rate due to the graviton
mass in terms of $a_{\rm max}$, which is the largest radius of the
bubble in Euclidean time
(Eqs.~(\ref{eq:massa})--(\ref{eq:Bdefine})). It is found that in the
thin-wall approximation, $a_{\rm max}$ coincides with the scale
factor for the false vacuum, hence the contributions from the
graviton mass cancel with the counter term of the false vacuum.
Corrections to the CDL tunneling rate appear when one considers the
deviations from thin-wall approximation, and its value varies
monotonically with respect to $a_{\rm max}$ until the HM case, as
illustrated in Fig.~\ref{fig:Bmass}.

Moreover, provided with the same shape of potential which satisfies
the condition for the thin-wall approximation, we compare the
probabilities for the HM and CDL process. It is found that if the
typical value for the graviton mass is larger than the inverse of
bubble radius, the HM process may dominate over the CDL one, which
is very different from the situation in GR. Hence, in dRGT massive
gravity theory, not only the shape of the potential but also the
value of the parameters $\alpha_3$, $\alpha_4$ and $m_g$ will
qualitatively influence the tunneling process.

On the other hand, it is known that dRGT massive gravity theory
suffers from some
problems~\cite{DeFelice:2012mx,Acausality,Acausality1}. Hence, as
one step towards a more realistic model, it is necessary to study
the tunneling issues in the extended massive gravity theories: for
example, quasi-dilaton massive
gravity~\cite{DeFelice:2013,Quasidilaton:2013,Quasidilaton:2013perturb,GHLMT,FM:2013a,FGM:2013},
varying-mass massive
gravity~\cite{Huang:2012,Huang:2013,Saridakis:2012,LSS:2013,HST:2013,WPC:2013,BHNMS:2013}
or $SO(3)$ massive gravity~\cite{lin:2013a,lin:2013b}. Especially,
in the varying-mass massive gravity theory, due to the mass
dependence, the effective cosmological constant $\Lambda_\pm$ of
some vacuum may become larger even with a relatively smaller
potential energy, which may imply a scenario of tunneling from a
lower potential energy to a higher one. Investigations of the
corresponding tunneling process is one of the future studies.

\appendix

%%%%%%%%%%%%%%
\section{Calculation of perturbations of action around HM solution}\label{perturbHM}
%%%%%%%%%%%%%%
In this appendix, we present detailed calculations of perturbations
in Euclidean action around HM solutions. It is obvious that from  in
Eq.~(\ref{dsformal}), provided with Eq.~(\ref{eq:perturba}), for our
calculation of the perturbations of the action up to 2nd order, we
should firstly consider the perturbations in the term $\sqrt{-
(f^\prime)^2}$. Under perturbation $f\longrightarrow f_0+\delta f$,
we have:
\begin{eqnarray}\label{eq:perterbf}
\sqrt{- (f^\prime)^2}=\sqrt{- (f_0^\prime)^2}~\left|1+\frac{(\delta
f)^\prime}{f_0^\prime}\right|
\end{eqnarray}

The constraint Eq.~(\ref{eq:II}) implies the relationship $\delta
f=X_\pm\delta a/b_{,f}$, so we have
\begin{eqnarray}
(\delta f)^\prime=\frac{X_\pm}{b_{,f}}(\delta a)^\prime-
\frac{X_\pm}{ b_{,f}^2}\left(b_{,f}\right)^\prime\delta a\,,
\end{eqnarray}
while
\begin{eqnarray}
b^\prime_0=b_{,f}f_0^\prime=X_\pm
a_0^\prime\quad\Longrightarrow\quad f_0^\prime=\frac{X_\pm
a_0^\prime}{b_{,f}}\,.
\end{eqnarray}

Combining these two equations, we then obtain the following
relationship:
\begin{eqnarray}\label{f}
\frac{(\delta
f)^\prime}{f_0^\prime}=\frac{1}{a_0^\prime}\left((\delta
a)^\prime-\frac{\left(b_{,f}\right)^\prime}{b_{,f}}\delta
a\right)\,.
\end{eqnarray}

On the other hand, noting that
$b_{,f}=\pm\sqrt{\tilde{\alpha}^2\cos^2\left(\tilde{H}_{\rm
HM}\tau\right)-1}$ with $\tilde{\alpha}\equiv FX_\pm/\tilde{H}_{\rm
HM}$, we have:
\begin{eqnarray}
\left(b_{,f}\right)^\prime=\pm\frac{-\tilde{\alpha}^2\tilde{H}_{\rm
HM}\cos\left(\tilde{H}_{\rm HM}\tau\right)\sin\left(\tilde{H}_{\rm
HM}\tau\right)}{\sqrt{\tilde{\alpha}^2\cos\left(\tilde{H}_{\rm
HM}\tau\right)-1}}\,,
\end{eqnarray}
from which one immediately obtains that
\begin{eqnarray}\label{bf}
\frac{\left(b_{,f}\right)^\prime}{b_{,f}}=\frac{\tilde{\alpha}^2\tilde{H}_{\rm
HM}\cos\left(\tilde{H}_{\rm HM}\tau\right)\sin\left(\tilde{H}_{\rm
HM}\tau\right)}{1-\tilde{\alpha}^2\cos^2\left(\tilde{H}_{\rm
HM}\tau\right)}\,.
\end{eqnarray}

Inserting Eqs.~(\ref{eq:perturba}) and (\ref{bf}) into
Eq.~(\ref{f}), we obtain the following expression:
\begin{eqnarray}\label{eq:fcorr}
\frac{(\delta f)^\prime}{f_0^\prime}=\frac{\varepsilon_M^2{H}_{\rm
HM}^2\cos^2\left(\tilde{H}_{\rm
HM}\tau\right)}{8}\left(3+\frac{\tilde{\alpha}^2\cos^2\left(\tilde{H}_{\rm
HM}\tau\right)}{1-\tilde{\alpha}^2\cos^2\left(\tilde{H}_{\rm
HM}\tau\right)}\right)>0\,.
\end{eqnarray}

Hence, from Eqs.~(\ref{eq:perterbf}) and (\ref{eq:fcorr}), we
conclude that at range $\tilde{H}_{\rm HM}\tau\in(0, \pi/2)$, the
the second order perturbation arising from the term $\sqrt{-
(f^\prime)^2}$ can be expressed in the following way:
\begin{eqnarray}\label{df}
\delta\sqrt{- (f^\prime)^2}=\sqrt{-(f_0^\prime)^2}~\frac{(\delta
 f)^\prime}{f_0^\prime}=\frac{\varepsilon_M^2X_\pm\tilde{H}_{\rm
HM}^2\sin\left(\tilde{H}_{\rm
HM}\tau\right)\cos^2\left(\tilde{H}_{\rm
HM}\tau\right)}{8\sqrt{1-\tilde{\alpha}^2\cos^2\left(\tilde{H}_{\rm
HM}\tau\right)}}\left[3+\frac{\tilde{\alpha}^2\cos^2\left(\tilde{H}_{\rm
HM}\tau\right)}{1-\tilde{\alpha}^2\cos^2\left(\tilde{H}_{\rm
HM}\tau\right)}\right]\,.
\end{eqnarray}

Now let us calculate the two parts in Eq.~(\ref{dsformal})
separately. The first part reads
\begin{eqnarray}\label{p1}
3\int_0^{\pi/2\tilde{H}_{\rm HM}} {\rm d}\tau~a_{\rm HM}^2\sqrt{-
(f^\prime)^2}\delta a&=&\frac{3}{\tilde{H}_{\rm
HM}^2}\int_0^{\pi/2\tilde{H}_{\rm HM}} {\rm
d}\tau~\cos^2\left(\tilde{H}_{\rm
HM}\tau\right)\frac{X_\pm\sin\left(\tilde{H}_{\rm
HM}\tau\right)}{\sqrt{1-\tilde{\alpha}^2\cos^2\left(\tilde{H}_{\rm
HM}\tau\right)}}\frac{\varepsilon_M^2H_{\rm HM}^2}{8\tilde{H}_{\rm
HM}}\cos^3\left(\tilde{H}_{\rm HM}\tau\right)\nonumber\\
&=&\frac{3\varepsilon_M^2H_{\rm HM}^2X_\pm}{8\tilde{H}_{\rm
HM}^4}\int_0^{\pi/2} {\rm
d}z~\cos^5\left(z\right)\frac{\sin\left(z\right)}{\sqrt{1-\tilde{\alpha}^2\cos^2\left(z\right)}}\nonumber\\
&=&-\frac{3\varepsilon_M^2H_{\rm HM}^2X_\pm}{8\tilde{H}_{\rm
HM}^4}\int_1^{0} {\rm
d}s~\frac{s^5}{\sqrt{1-\tilde{\alpha}^2s^2}}\nonumber\\
&=&-\frac{3\varepsilon_M^2H_{\rm HM}^2X_\pm}{8\tilde{H}_{\rm
HM}^4}\left[\frac{-8+\sqrt{1-\tilde{\alpha}^2}\left(8+4\tilde{\alpha}^2+3\tilde{\alpha}^4\right)}{15\tilde{\alpha}^6}\right]\,,
\end{eqnarray}
where in the second step, $z=\tilde{H}_{\rm HM}\tau$, and in the
third step, $s=\cos(z)$. Meanwhile, the second part of
Eq.~(\ref{dsformal}) can be calculated by using Eq.~(\ref{df}) as
follows:
\begin{eqnarray}\label{p2}
\int_0^{\pi/2\tilde{H}_{\rm HM}} {\rm d}\tau~a_{\rm
HM}^3\delta\sqrt{- (f^\prime)^2}&=&\frac{\varepsilon_M^2{H}_{\rm
HM}^2}{\tilde{H}_{\rm HM}^3}\int_0^{\pi/2\tilde{H}_{\rm HM}} {\rm
d}\tau~\frac{X_\pm\sin\left(\tilde{H}_{\rm
HM}\tau\right)\cos^5\left(\tilde{H}_{\rm
HM}\tau\right)}{8\sqrt{1-\tilde{\alpha}^2\cos^2\left(\tilde{H}_{\rm
HM}\tau\right)}}\left(3+\frac{\tilde{\alpha}^2\cos^2\left(\tilde{H}_{\rm
HM}\tau\right)}{1-\tilde{\alpha}^2\cos^2\left(\tilde{H}_{\rm
HM}\tau\right)}\right)\nonumber\\
&=&\frac{\varepsilon_M^2X_\pm{H}_{\rm HM}^2}{\tilde{H}_{\rm
HM}^4}\int_0^{\pi/2} {\rm
d}z~\frac{\sin\left(z\right)\cos^5\left(z\right)}{8\sqrt{1-\tilde{\alpha}^2\cos^2\left(z\right)}}
\left(3+\frac{\tilde{\alpha}^2\cos^2\left(z\right)}{1-\tilde{\alpha}^2\cos^2\left(z\right)}\right)\nonumber\\
&=&-\frac{\varepsilon_M^2X_\pm{H}_{\rm HM}^2}{8\tilde{H}_{\rm
HM}^4}\int_1^{0} {\rm d}s~\frac{s^5}{\sqrt{1-\tilde{\alpha}^2s^2}}
\left(3+\frac{\tilde{\alpha}^2s^2}{1-\tilde{\alpha}^2s^2}\right)\nonumber\\
&=&-\frac{\varepsilon_M^2X_\pm{H}_{\rm HM}^2}{8\tilde{H}_{\rm
HM}^4}\frac{1}{5\tilde{\alpha}^6}\left(8+\frac{-8+4\tilde{\alpha}^2+\tilde{\alpha}^4-2\tilde{\alpha}^6}{\sqrt{1-\tilde{\alpha}^2}}\right)\,.
\end{eqnarray}

Thus, inserting Eqs.~(\ref{p1}) and (\ref{p2}) into
(\ref{dsformal}), one finally obtains the second order perturbation
in Euclidean action:
\begin{eqnarray}\label{s2app}
\delta^{(2)}S =\frac{\pi^2m_g^2X_\pm Y_\pm{H}_{\rm
HM}^2\varepsilon_M^2}{2\tilde{H}_{\rm
HM}^4\sqrt{1-\tilde{\alpha}^2}}\,.
\end{eqnarray}

%%%%%%%%%%%
\section{Rate of the CDL process to HM one in thin-wall approximation}\label{cdltoHM}
%%%%%%%%%%%
In this appendix, we derive Eq.~(\ref{Bcdlhm}) in details. For
simplicity, let us assume $K=1$ and set the effective cosmological
constant of the true vacuum $\Lambda_{\pm, \rm T}\equiv
V(\sigma_{\rm T})+\Lambda_\pm=0$, while that of the false vacuum
$\Lambda_{\pm, \rm F}\equiv V(\sigma_{\rm F})+\Lambda_\pm=\epsilon
M_{\rm Pl}^4\ll M_{\rm Pl}^4$. Then from Eq.~(\ref{Ball}), one
obtains:
\begin{align}\label{Bcomp}
B &\simeq-2\pi^2\left\{6\int^{a_0}_{0}a{\rm
d}a\left[1-\sqrt{K-(aH_{\rm F})^2}\right]-
a^3_0\Sigma\right\}\nonumber\\
&=-12\pi^2\left\{\frac{a_0^2}{2}+\frac{1}{\epsilon}\left[\left(1-\frac{a_0^2}{3}\epsilon\right)^\frac{3}{2}-1\right]\right\}
+2\pi^2a_0^3\Sigma\,,
\end{align}
which is stationary when
\begin{align}\label{a0extreme}
a_0=\frac{12\Sigma}{3\Sigma^2+4\epsilon}\,.
\end{align}
Inserting Eq.~(\ref{a0extreme}) into (\ref{Bcomp}), one obtains:
\begin{align}\label{Babso}
B\simeq12\pi^2\left[\frac{72\Sigma^2(\Sigma^2-4\epsilon)}{(3\Sigma^2+4\epsilon)^3}+\frac{1}{\epsilon}\left(1-\left|\frac{3\Sigma^2-4\epsilon}{3\Sigma^2+4\epsilon}\right|^3\right)\right]\,.
\end{align}

In order to evaluate the absolute value term in the right-hand side
of Eq.~(\ref{Babso}), we note that from Eq.~(\ref{eq:Bcri}), the
extreme value for tension $\Sigma$ can be also evaluated as follows:
\begin{align}\label{eq:Bcriapp}
\Sigma=\frac{2}{a_0}\left(1-\sqrt{1-\frac{a_0^2\epsilon}{3}}\right)\approx\frac{a_0\epsilon}{3}\,.
\end{align}

Since $a_0<H_{\rm F}^{-1}=\sqrt{3/\epsilon}$, Eq.~(\ref{eq:Bcriapp})
implies that $3\Sigma^2-4\epsilon<-3\epsilon<0$. Thus, from
Eq.~(\ref{Babso}), we finally obtain that:
\begin{align}\label{Bthinfull}
B&\simeq12\pi^2\left\{\frac{72\Sigma^2(\Sigma^2-4\epsilon)}{(3\Sigma^2+4\epsilon)^3}+\frac{1}{\epsilon}\left[1+\left(\frac{3\Sigma^2-4\epsilon}{3\Sigma^2+4\epsilon}\right)^3\right]\right\}\nonumber\\
&=\frac{24\pi^2}{\epsilon}\left(1+\frac{4\epsilon}{3\Sigma^2}\right)^{-2}\,.
\end{align}

Hence, the corresponding tunneling rate defined in
Eq.~(\ref{eq:rate}) is expressed as:
\begin{align}\label{Bcdlfull}
P_{\rm
thin-wall}\approx\exp\left[-\frac{24\pi^2}{\epsilon}\left(1+\frac{4\epsilon}{3\Sigma^2}\right)^{-2}\right]
\approx\exp\left(-\frac{24\pi^2}{\epsilon}+\frac{64\pi^2}{\Sigma^2}\right)\,,
\end{align}
while by using Eqs.~(\ref{eq:Bredefine}) and (\ref{eq:Bredefine1}),
that of thick-wall approximation can be expressed as~\cite{kklt}
\begin{align}\label{Bhmfull}
P_{\rm
thick-wall}&\approx\exp\left[-\frac{24\pi^2}{\epsilon}\left(1-\frac{\epsilon}{\Lambda_{\pm,
\rm eff}}\right)+\frac{4\pi^2m_g^2Y_\pm\mathfrak{B}^{\rm
mass}}{3F^4X_\pm^3}\right]\nonumber\\
&\approx\exp\left(-\frac{24\pi^2}{\epsilon}+\frac{24\pi^2}{\Lambda_{\pm,
\rm eff}}+\frac{4\pi^2m_g^2Y_\pm\mathfrak{B}^{\rm
mass}}{3F^4X_\pm^3}\right)\,.
\end{align}

Combining Eqs.~(\ref{Bcdlfull}) and (\ref{Bhmfull}), one can compare
the probability of the CDL process to that of HM as follows:
\begin{align}\label{Bcdlhmapp1}
\ln\left(\frac{P_{\rm CDL}}{P_{\rm
HM}}\right)=4\pi^2\left(\frac{16}{\Sigma^2}-\frac{6}{\Lambda_{\pm,
\rm eff}}-\frac{m_g^2Y_\pm\mathfrak{B}^{\rm mass}(a_{\rm
max}=H^{-1}_{\rm HM})}{3F^4X_\pm^3}\right)\,.
\end{align}

Noting that from Eq.~(\ref{tension}), one can make a comparison
between $\Sigma^2$ and $\Lambda_{\pm, \rm eff}$ as follows (here
$M_{\rm Pl}$ is recovered for comparison)
\begin{align}\label{tensionapp}
\frac{\Sigma}{M_{\rm Pl}\sqrt{\Lambda_{\pm, \rm
eff}}}&\sim\frac{\int_{\sigma_{\rm T}}^{\sigma_{\rm F}}{\rm
d}\sigma\sqrt{V(\sigma)-V(\sigma_{\rm
    T})}}{M_{\rm Pl}H_{\rm HM}}<\frac{\sqrt{V(\sigma_{\rm HM})-V(\sigma_{\rm T})}}{M_{\rm Pl}H_{\rm HM}}\Delta\sigma\nonumber\\
    &=\frac{\sqrt{V(\sigma_{\rm HM})+\Lambda_\pm}}{M_{\rm Pl}H_{\rm HM}}\Delta\sigma
    =\frac{\Delta\sigma}{M_{\rm Pl}}\sim\frac{\big|\sigma_{\rm HM}-\sigma_{\rm T}\big|}{M_{\rm Pl}}\,,
\end{align}
where $\Delta\sigma\equiv|\sigma_{\rm F}-\sigma_{\rm T}|$ and we
have used $V(\sigma_{\rm T})=-\Lambda_\pm$. On the other hand,
expending the potential near $\sigma_{\rm HM}$, we have
\begin{align}\label{potentialHM}
V(\sigma)=V(\sigma_{\rm HM})-\frac{M^2}{2}(\sigma-\sigma_{\rm
HM})^2+...\,,
\end{align}
where $M^2 \equiv -{\rm d}^2V(\sigma)/{\rm
d}\sigma^2|_{\sigma=\sigma_{\rm HM}}$. So one obtains
\begin{align}\label{T-HM}
\frac{M^2}{2}(\sigma_{\rm T}-\sigma_{\rm HM})^2\approx V(\sigma_{\rm
HM})-V(\sigma_{\rm T})<H_{\rm
HM}^2\quad\Longrightarrow\quad\big|\sigma_{\rm T}-\sigma_{\rm
HM}\big|<\frac{H_{\rm HM}}{M}\,.
\end{align}

Inserting this into Eq.(\ref{potentialHM}), we find that
\begin{align}\label{compar1}
\frac{\Sigma}{M_{\rm Pl}\sqrt{\Lambda_{\pm, \rm eff}}}<\frac{H_{\rm
HM}}{M}\,.
\end{align}

In the thin-wall approximation, the typical height of the local
maximum of the potential is much larger than its width so that
$M^2\gg H_{\rm HM}^2$, then one obtains that
\begin{align}\label{comparresult}
\Sigma^2\ll M_{\rm Pl}^2\Lambda_{\pm, \rm eff}\,.
\end{align}

We note that Eq.~(\ref{comparresult}) can be justified in another
way: from Eq.~(\ref{eq:Bcri}), using the relationship $H_{\rm
F}^2-H_{\rm T}^2=\epsilon/3$, the radius of the bubble can be
expressed as~\cite{Coleman:1980}
\begin{align}\label{aradius}
a_0=\frac{4\Sigma}{\sqrt{\left(\Sigma^2+\frac{4\epsilon}{3}\right)^2+16\Sigma^2H_{\rm
T}^2}}=\frac{12\Sigma}{3\Sigma^2+4\epsilon}\sim\frac{1}{\Sigma}\,,
\end{align}
where we have used the assumption $H_{\rm T}^2\equiv\Lambda_{\pm,
\rm T}/3=0$ and $\Sigma^2\gtrsim \mathcal{O}(\epsilon)$. On the
other hand, using Eqs.~(\ref{eq:adot}) and (\ref{eq:derisig}), one
finds that
\begin{align}\label{aderiva}
a^{\prime}=\sqrt{1-a^2H_{\rm T}^2}=1\,,
\end{align}
so using the relationship ${\rm d}\sigma=\sigma^\prime{\rm
d}a/a^\prime$, the thickness of the wall can be approximately
evaluated as
\begin{align}\label{thickness}
\Delta
a=\frac{a^\prime\Delta\sigma}{\sigma^\prime}\approx\frac{\Sigma}{V(\sigma_{\rm
HM})-V(\sigma_{\rm T})}=\frac{\Sigma}{\Lambda_{\pm, \rm eff}}\,,
\end{align}
where $\Delta\sigma\equiv|\sigma_{\rm F}-\sigma_{\rm
T}|\approx\Sigma/\sqrt{V-V(\sigma_{\rm T})}$. The thin-wall
approximation is valid if $a_0/\Delta a\gg1$, so one obtains
$\Sigma^2\ll\Lambda_{\pm, \rm eff}$~\cite{kklt}, which verifies
Eq.~(\ref{comparresult}). Thus, in the thin-wall approximation,
Eq.~(\ref{Bcdlhmapp1}) reduces to the following form:
\begin{align}\label{Bcdlhmapp}
\ln\left(\frac{P_{\rm CDL}}{P_{\rm
HM}}\right)&\approx4\pi^2\left(\frac{16}{\Sigma^2}-\frac{m_g^2Y_\pm\mathfrak{B}^{\rm
mass}(a_{\rm max}=H^{-1}_{\rm
HM})}{3F^4X_\pm^3}\right)\nonumber\\
&=4\pi^2\left[\frac{16}{\Sigma^2}-\frac{m_g^2X_\pm Y_\pm }{3}\left(
\frac{A(\alpha_{\rm F})}{H_{\rm F}^4}-\frac{A(\alpha_{\rm
HM})}{H_{\rm HM}^4}\right)\right]\,,
\end{align}
where we used the Eq.~(\ref{eq:A}) for definition of function
$A(\alpha)$. In the context of GR where $m_g=0$, provided that the
CDL instantons exist, the CDL process always dominates over the HM
one~\cite{kklt}.

\begin{acknowledgments}
We thank Stefano Ansoldi, Qing-guo Huang and Kazuyuki Sugimura for
helpful discussions. This work was supported in part by the
Grant-in-Aid for the Global COE Program ``The Next Generation of
Physics, Spun from Universality and Emergence'' from the Ministry of
Education, Culture, Sports, Science and Technology (MEXT) of Japan,
and by JSPS Grant-in-Aid for Scientific Research (A) No.~21244033.
RS is supported by a JSPS Grant-in-Aid through the JSPS postdoctoral
fellowship No. 23$\cdot$3430.
\end{acknowledgments}

\end{document}